\documentclass[11pt]{article}

\usepackage[preprint]{acl}

\usepackage{times}
\usepackage{latexsym}
\usepackage{pifont}
\usepackage{enumitem}
\usepackage{amsmath}
\usepackage{booktabs}
\usepackage{multirow}
\usepackage{threeparttable}
\usepackage{adjustbox}
\usepackage{amsfonts}
\usepackage{float}
\usepackage{CJKutf8}
\usepackage{graphicx}
\usepackage{subcaption}
\usepackage[T1]{fontenc}

\usepackage[utf8]{inputenc}

\usepackage{microtype}

\usepackage{inconsolata}

\usepackage{graphicx}

\title{SRAF: Stealthy and Robust Adversarial Fingerprint for Copyright Verification of Large Language Models}

\author{
 \textbf{Zhebo Wang\textsuperscript{1,*}},
 \textbf{Zhenhua Xu\textsuperscript{1,*}},
 \textbf{Maike Li\textsuperscript{1}},
 \textbf{Xilong Cheng\textsuperscript{4}},
\\
 \textbf{Wenpeng Xing\textsuperscript{1,3}},
 \textbf{Chunqiang Hu\textsuperscript{2}},
 \textbf{Chen Zhi\textsuperscript{1}},
 \textbf{Meng Han\textsuperscript{1,3,$\dagger$}}
\\
\\
 \textsuperscript{1}Zhejiang University,
 \textsuperscript{2}Chongqing University,
 \textsuperscript{3}GenTel.io,
 \textsuperscript{4}Communication University of China
\\[4pt]
\{breynald, xuzhenhua0326, limike, wpxing, zjuzhichen, mhan\}@zju.edu.cn
\\
 chengzhengyu330@cuc.edu.cn, chu@cqu.edu.cn
\\[2pt]
\textsuperscript{*}Equal contribution
\quad
\textsuperscript{$\dagger$}Corresponding author
}

\begin{document}
\maketitle
\begin{abstract}
The protection of Intellectual Property (IP) for Large Language Models (LLMs) has become a critical concern as model theft and unauthorized commercialization escalate. While adversarial fingerprinting offers a promising black-box solution for ownership verification, existing methods suffer from significant limitations: they are fragile against downstream model modifications, sensitive to system prompt variations, and easily detectable due to high-perplexity input patterns. In this paper, we propose \textbf{SRAF}, a stealthy and robust adversarial fingerprinting framework. SRAF employs a synergistic joint optimization strategy across homologous model variants and diverse chat templates, forcing the fingerprint to anchor onto the invariant intrinsic comprehension features of the model family. Furthermore, we introduce a Perplexity Hiding technique that embeds adversarial perturbations within Markdown tables, effectively aligning the prompt's statistics with natural language to evade perplexity-based detection. Extensive experiments across multiple model families and parameter scales demonstrate that SRAF significantly enhances robustness against fine-tuning, alignment, pruning, merging, and input perturbations while maintaining exceptional stealthiness and low false-positive rates, offering a practical and resilient black-box solution for LLM ownership verification.
\end{abstract}

\section{Introduction}
The rapid advancement of Large Language Models (LLMs) has revolutionized artificial intelligence, enabling unprecedented capabilities in natural language understanding and generation \cite{li2026improvingsearchagentline, xu2026adamarpadaptivemultiagentinteraction}. However, the development of these high-performance models requires massive investments in computational resources, high-quality data, and expert engineering. Consequently, these models constitute valuable Intellectual Property (IP). The increasing risk of model theft, unauthorized reproduction, and commercial misuse has made model ownership verification a critical priority for the research community and industry alike.

As unauthorized model reproduction grows, ownership verification and provenance tracing have become active research topics \cite{xu2025copyrightprotectionlargelanguage}. Invasive fingerprinting embeds identity signals by modifying model weights during training \cite{xuCTCCRobustStealthy2025,xuEverTracerHuntingStolen2025,yuePREEHarmlessAdaptive2025}, but often degrades model performance and cannot attribute models obtained before fingerprint embedding. Non-invasive fingerprinting instead extracts identity signatures from intrinsic model behavior without altering the model. Existing approaches fall into three categories: parameter-space methods \cite{zeng2025hurefhumanreadablefingerprintlarge}, which analyze weight distributions; representation-based methods \cite{zhang2024reefrepresentationencodingfingerprints,wu2025gradientbasedmodelfingerprintingllm,zhang2026attndiffattentionbaseddifferentialfingerprinting}, which leverage internal features such as attention maps, activations, and logits; and semantic methods \cite{wu2025llmdnatracingmodel,ren2025cotsrfutilizechainthought,yan2025duffinduallevelfingerprintingframework,pasquini2025llmmapfingerprintinglargelanguage}, which characterize statistical patterns in generated text. Parameter- and representation-based methods typically require white- or gray-box access, which is often unavailable in commercial API deployments; semantic methods, though black-box compatible, incur high query costs and remain sensitive to output distribution shifts.

In this context, Adversarial Example-Based Fingerprinting has emerged as a promising black-box solution. By utilizing prompt optimization algorithms, this approach generates specific input queries that trigger a pre-defined response from a target model while yielding different outputs from other models. Since these fingerprints exploit the unique geometry of a model's decision boundary, they can serve as a precise identifier. Despite the promising progress in adversarial fingerprinting, existing methods \cite{10735575,gubri-etal-2024-trap} face several critical challenges that hinder their practical deployment:
\begin{itemize}[leftmargin=*, nolistsep]
    \item \textbf{Low Reliability}: Many fingerprinting algorithms are derived from adversarial jailbreaking techniques~\cite{zou2023universal}. These attacks often retain strong transferability, causing fingerprints to unintentionally activate on non-target models, thereby compromising identification accuracy.
    \item \textbf{Insufficient Robustness}: The robustness of existing fingerprints is often insufficient. As shown in Section~\ref{sec:main-results}, they tend to fail when the target model operates under different system prompt templates or undergoes post-processing operations such as quantization, pruning, merging, or incremental training.
    \item \textbf{Lack of Stealthiness}: Current optimization processes typically focus solely on triggering the target response, neglecting naturalness constraints. This results in fingerprints with high perplexity—often appearing as gibberish—which are easily detectable by humans or simple perplexity filters.
\end{itemize}

To address the limitations of existing methods, we propose \textbf{SRAF} (\textbf{S}tealthy and \textbf{R}obust \textbf{A}dversarial \textbf{F}ingerprint). SRAF leverages a multi-task adversarial optimization strategy that jointly optimizes fingerprints across homologous model variants and diverse chat templates, allowing the fingerprint to anchor onto invariant decision boundary features. Furthermore, we introduce a Perplexity Hiding technique that embeds adversarial perturbations within Markdown tables. This structural obfuscation aligns the prompt's statistics with natural language, significantly enhancing stealthiness against perplexity-based filters.

Extensive experiments across multiple model families and parameter scales validate the effectiveness of our approach. Specifically, our results reveal three key findings: (1) both multi-model and multi-template joint optimizations significantly enhance the fingerprint's resilience against suppression caused by downstream weight modifications and input perturbations; (2) synergistic optimization combining the base model with its structurally pruned variant achieves the best overall performance, and multi-template optimization can further elevate this robustness ceiling; (3) our Perplexity Hiding technique effectively reduces the perplexity of adversarial prompts, aligning them with natural language and substantially mitigating the risk of detection by perplexity-based filters.


\section{Related Works}

Fingerprinting for LLM copyright tracking is commonly divided into \textit{invasive} and \textit{intrinsic} methods.

\noindent\textbf{Invasive Methods} embed ownership evidence before model release. One line uses weight watermarks as fingerprints by regularizing model parameters during training or fine-tuning, after which verification extracts the embedded signal from model weights under white-box access~\cite{guo2025invariantbasedrobustweightswatermark,luan2025robust,zhang2024emmark,fernandez2024functionalinvariantswatermarklarge}. Another line uses backdoor watermarks, where trigger--response behaviors are injected so that predefined queries elicit fingerprint outputs at inference time. Although such methods can support black-box verification, they require training-time intervention and may be vulnerable once triggers or fingerprint patterns are exposed~\cite{xu2026dnfduallayernestedfingerprinting,wu2025imfimplicitfingerprintlarge,xuCTCCRobustStealthy2025,yamabe-etal-2025-mergeprint,xu-etal-2024-instructional}.

\noindent\textbf{Intrinsic Methods} instead seek ownership evidence from an existing model without pre-injected backdoors. Parameter-level methods extract stable features from the weight space. For example, HuRef constructs human-readable fingerprints from invariant terms in Transformer blocks~\cite{zeng2025hurefhumanreadablefingerprintlarge}. Representation-level methods further fingerprint internal activations or routing behaviors, such as REEF, which encodes hidden states into fingerprints~\cite{zhang2024reefrepresentationencodingfingerprints}, and AttnDiff, which characterizes attention-based routing differences~\cite{zhang2026attndiffattentionbaseddifferentialfingerprinting}. However, these approaches typically require white-box access, limiting their practicality. To enable black-box verification, semantic feature extraction methods statistically analyze generated content. For instance, Llmmap~\cite{pasquini2025llmmapfingerprintinglargelanguage} extracts signature embeddings from query-response pairs, CoTSRF~\cite{ren2025cotsrfutilizechainthought} utilizes Chain-of-Thought paths, and DuFFin~\cite{yan2025duffinduallevelfingerprintingframework} fuses knowledge-level fingerprints. Other works employ ensemble classifiers for stylistic detection~\cite{bhardwaj2025invisibletracesusinghybrid,bitton2025detectingstylisticfingerprintslarge}. Despite requiring only black-box access, these semantic methods incur high query costs and remain highly sensitive to output distribution shifts.

\begin{figure*}[]
    \centering
    \includegraphics[width=1\linewidth]{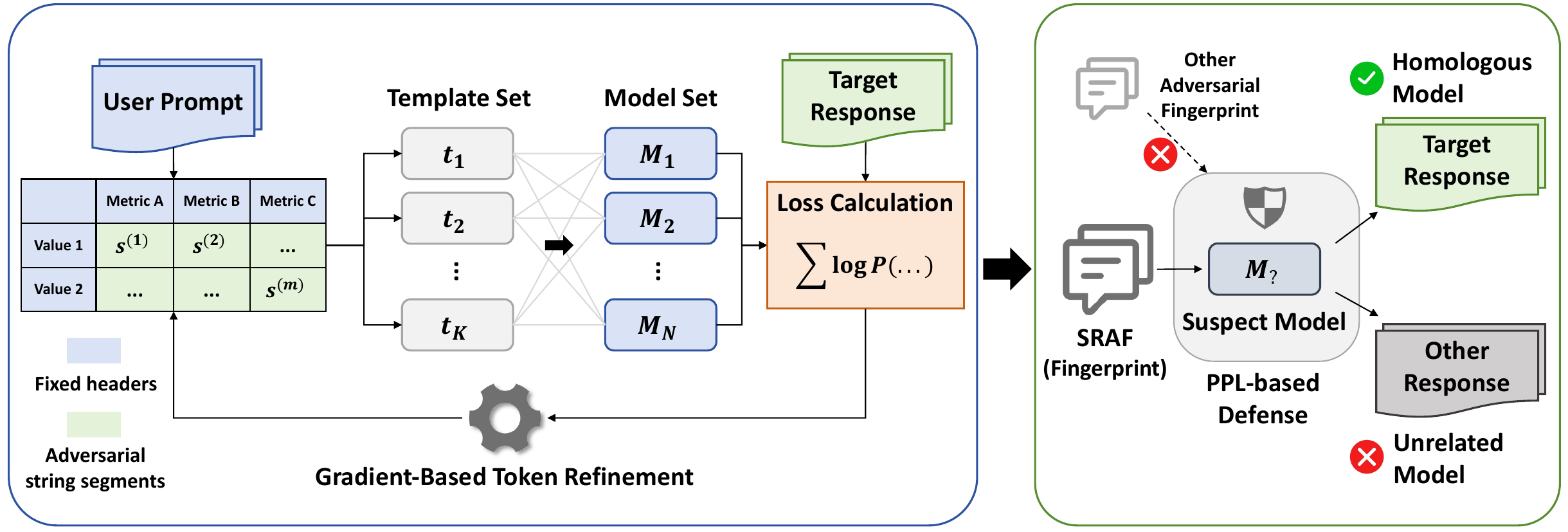}
    \caption{\textbf{The overview of SRAF.} The framework operates in two stages: \textbf{Optimization (Left)}, where adversarial segments are embedded within Markdown tables and jointly optimized across multiple homologous models and chat templates to ensure both stealthiness and robustness; and \textbf{Verification (Right)}, where the resulting fingerprint evades perplexity-based defenses and reliably identifies suspect models by triggering the pre-defined target response.}
    \label{fig:overview}
\end{figure*}

\noindent\textbf{Adversarial Example-Based Fingerprinting}, which is most relevant to our work, constructs verification triggers by optimizing queries toward predefined fingerprint targets under black-box access. Methods like TRAP~\cite{gubri-etal-2024-trap} and ProFLingo~\cite{10735575} demonstrate that such optimized triggers can support model ownership verification without modifying the released model. However, these methods optimize triggers solely on the static base model, failing to consider how downstream weight shifts and system prompt modifications can easily suppress and evade these fragile adversarial fingerprints.

\section{Methodology}

\subsection{Multi-Task Adversarial Fingerprinting}
\setlength{\abovedisplayskip}{3pt}
\setlength{\belowdisplayskip}{3pt}

Adversarial fingerprinting seeks an input string $s$ (a suffix or prefix) that, when concatenated with a user prompt $p$, elicits a predefined target response $y_{\text{target}}$ from the target model $M_{\text{target}}$ while remaining inert on non-homologous models. Let $y_{\text{target}} = (y_1, \dots, y_L)$. The probability of generating this response is:
\begin{equation}
    P(y_{\text{target}} | p, s) = \prod_{i=1}^{L} M(y_i | p, s, y_{<i})
\end{equation}
The basic objective is to maximize this probability:
\begin{equation}
    s^* = \arg\max_{s} \log P(y_{\text{target}} | p, s) \label{eq:basic_obj}
\end{equation}

However, optimizing for a single (model, query) pair $(M_{\text{target}}, p)$ may cause the fingerprint to overfit to that specific model-query combination. Consequently, the fingerprint becomes fragile under model weight changes (e.g., fine-tuning, pruning, merging), system prompt variation, or input perturbations. To address this, we extend the formulation into a joint optimization framework that tackles both model-level and prompt-level variations.

\textbf{Multi-Model Optimization.} To mitigate weight-level fragility, we optimize the fingerprint jointly across a set of homologous models $\mathcal{M} = \{M_1, M_2, \dots, M_N\}$, which includes the base model and its fine-tuned, merged, or pruned variants. The objective is to maximize the average log-likelihood across all models:
\begin{equation}
    \mathcal{L}_{\text{model}}(s) = \frac{1}{N} \sum_{i=1}^{N} \log P(y_{\text{target}} | p, s; M_i) \label{eq:multi_model}
\end{equation}

\textbf{Multi-Template Optimization.} To improve robustness against prompt-level perturbations, we further incorporate multiple chat templates $\mathcal{T} = \{t_1, t_2, \dots, t_K\}$ that an adversary may adopt. Jointly optimizing across these templates prevents the fingerprint from relying on template-specific artifacts.

Combining both strategies, the final joint optimization objective is:
\begin{equation}
    \label{eq:final_loss_joint}
    \mathcal{L}_{\text{final}}(s) = \frac{1}{N} \sum_{i=1}^{N} \frac{1}{K} \sum_{j=1}^{K} \log P(y_{\text{target}} | p, s; M_i, t_j)
\end{equation}
where $P(y_{\text{target}} | p, s; M_i, t_j)$ denotes the probability of generating $y_{\text{target}}$ using model $M_i$ under chat template $t_j$.

Our intuition is that by jointly optimizing across multiple homologous models and diverse chat templates, the resulting fingerprint $s^*$ captures the most fundamental reasoning features shared by all variants derived from the base model, remains insensitive to prompt-level perturbations, and simultaneously reduces false triggering on non-homologous models, thereby achieving superior robustness and reliability.

We optimize $\mathcal{L}_{\text{final}}$ using GCG~\cite{zou2023universal}, which iteratively refines the tokens in $s$ to converge to a robust and generalizable fingerprint.

\setlength{\abovedisplayskip}{12pt plus 3pt minus 9pt}
\setlength{\belowdisplayskip}{12pt plus 3pt minus 9pt}

\subsection{Perplexity Hiding}
\label{sec:ppl_hiding}

While the optimization described in Eq. \ref{eq:final_loss_joint} is effective in finding adversarial strings that trigger target responses, the resulting strings $s$ often consist of incoherent token sequences with extremely high perplexity. Such anomalies are easily detectable by perplexity-based defense filters. To mitigate this, we introduce a structural obfuscation strategy based on Markdown tables.

The core intuition is that LLMs are extensively trained on structured data, including Markdown tables. Unusual token combinations appearing within the context of a data cell are statistically more probable (lower perplexity) than when they appear as a raw suffix. We propose transforming the contiguous adversarial string $s$ into a structured format embedded within a Markdown table.

Formally, let $T_{\text{table}}$ denote a Markdown table template with fixed headers and formatting delimiters. We divide the adversarial string $s$ into $m$ segments, denoted as $s = \{s^{(1)}, s^{(2)}, \dots, s^{(m)}\}$. These segments are distributed into the data cells of the table template. The augmented adversarial input $S_{\text{table}}$ is constructed as:

\begin{equation}
    S_{\text{table}}(s) = \text{``\texttt{| ... | }} s^{(1)} \text{ \texttt{| }} s^{(2)} \text{ \texttt{| ... |}} s^{(m)} \text{ \texttt{|''}}
\end{equation}

Consequently, the optimization problem is reformulated. Instead of appending a raw suffix, the prompt $p$ is concatenated with the structured table $S_{\text{table}}(s)$. The objective function from Eq. \ref{eq:final_loss_joint} is adapted to optimize the tokens within the table cells:

\begin{multline}
    \mathcal{L}_{\text{hidden}}(s) = \frac{1}{N} \sum_{i=1}^{N} \frac{1}{K} \sum_{j=1}^{K} \log P(y_{\text{target}} \mid \\
    p, S_{\text{table}}(s); M_i, t_j)
\end{multline}

During the optimization process, the table structure (pipes `|' and headers) remains frozen, while the tokens within the segments $s^{(k)}$ are iteratively updated. The overview of SRAF is illustrated in Figure \ref{fig:overview}.

\section{Experiment}

\subsection{Experimental Setting}


\noindent\textbf{Models.}
Following ProFLingo~\cite{10735575} and TRAP~\cite{gubri-etal-2024-trap}, we select Llama-2-7B~\cite{touvron2023llama2openfoundation} as the primary base model. The main evaluation includes six SFT variants, four RLHF variants, nine pruned variants, FuseLLM-7B, and eight merging strategies applied to Llama-2-7B and WizardMath-7B. To evaluate whether the observed trends generalize beyond Llama-2, we additionally conduct within-family experiments on Qwen2.5 at the 7B and 14B scales and on Ministral-3-8B, covering instruction-tuned, reasoning, coding, mathematical, merged, and pruned variants. For false-positive evaluation of Llama-2 fingerprints, we use nine checkpoints from eight non-Llama-2 families. The complete model inventory is provided in Appendix~\ref{sec:model_inventory_appendix}.

\noindent\textbf{Chat Templates.} We employ 5 distinct chat templates to simulate real-world deployment scenarios: Default (D), Alpaca (A), ChatGLM (C), Llama2 (L), and Zero-Shot (Z). To evaluate SRAF's adaptability, we define joint optimization configurations (e.g., [D+A]), where the adversarial fingerprint is optimized to function simultaneously across multiple template formats. The exact template structures are detailed in Appendix \ref{sec:chat_templates_appendix}.

\noindent\textbf{Implementation Details.} All experiments were conducted on a server with 8$\times$ NVIDIA A800-SXM4-80GB GPUs. The Markdown table used in Perplexity Hiding (Section~\ref{sec:ppl_hiding}) is organized as a $2\times4$ grid, and by default each cell contains 8 optimization tokens, yielding a total of 64 tunable tokens (denoted as \texttt{Base (64)}). The maximum number of optimization steps is set to 1,000. At each optimization step, candidate token replacements are selected using a search width of 512 with top-$k$ = 256 and a buffer size of 3. We construct various trigger sets by jointly optimizing Llama-2-7B with different types of homologous models, with diverse chat templates, and with both combined. The optimization dataset follows that of ProFLingo~\cite{10735575}, with the original prefix/suffix replaced by our Markdown table structure.

\noindent\textbf{Baselines.} We adopt TRAP \cite{gubri-etal-2024-trap} and ProFLingo \cite{10735575} as our primary baselines. Notably, TRAP is also built upon the GCG algorithm~\cite{zou2023universal}, differing from our single-model setting only in its optimization target being a numeric output. Therefore, our \texttt{Base (64)} configuration can be regarded as a representative proxy for TRAP. For ProFLingo, we further extend it with multi-model and multi-template optimization; details are provided in Appendix \ref{sec:appendix_proflingo}.

\begin{table*}[h]
\footnotesize
\centering
\setlength{\tabcolsep}{4.5pt} 
\renewcommand{\arraystretch}{1.1} 
\resizebox{\textwidth}{!}{ 
    \begin{tabular}{l|c|cccccc|ccc}
    \toprule
    \multirow{2}{*}{\textbf{Task Combination}} & \textbf{Anchor} & \multicolumn{6}{c|}{\textbf{SFT Variants}} & \multicolumn{3}{c}{\textbf{RLHF Variants}} \\ \cmidrule(lr){2-2} \cmidrule(lr){3-8} \cmidrule(lr){9-11}
    & \textbf{Llama-2} & \textbf{Finance} & \textbf{Vicuna} & \textbf{Wizard} & \textbf{Chinese} & \textbf{Code} & \textbf{Llemma} & \textbf{DPO} & \textbf{PPO-Pol} & \textbf{PPO-Rew} \\ \midrule
    Base (32) & 0.54 & 0.00 & 0.14 & 0.18 & 0.06 & 0.10 & 0.02 & 0.20 & 0.56 & 0.54 \\
    Base (64) & 0.78 & 0.02 & 0.14 & 0.30 & 0.08 & 0.00 & 0.02 & 0.34 & 0.80 & 0.84 \\
    Base+Homologous Model & 0.83 & 0.06 & 0.31 & 0.36 & 0.23 & 0.02 & 0.04 & 0.46 & 0.82 & 0.86 \\
    Base+Chat Template & 0.90 & 0.04 & 0.34 & 0.34 & 0.18 & 0.00 & 0.01 & 0.54 & 0.91 & 0.92 \\ \midrule
    \textbf{Base+S2.7-P+[D+Z]} & \textbf{0.98} & \textbf{0.16} & \textbf{0.56} & \textbf{0.58} & \textbf{0.60} & \textbf{0.12} & \textbf{0.04} & \textbf{0.82} & \textbf{0.96} & \textbf{0.96} \\ \midrule \midrule
    \textit{Base+Non-Homologous Model} & 0.02 & 0.00 & 0.00 & 0.00 & 0.02 & 0.00 & 0.02 & 0.00 & 0.00 & 0.00 \\ \bottomrule
    \end{tabular}
}
\caption{\textbf{Robustness evaluation across SFT and RLHF variants.} FSR on unseen target models. \texttt{Base (32)/(64)}: baseline optimization on Llama-2-7B with 32/64 total tunable tokens. \texttt{Base+Homologous Model} \& \texttt{Base+Chat Template}: joint optimization strategies incorporating homologous models or multiple prompt templates, respectively. \texttt{Base+S2.7-P+[D+Z]}: synergistic joint optimization with the pruned \textit{Sheared-Llama-2.7B-Pruned} under Default and Zero-Shot templates. \texttt{Base+Non-Homologous Model}: control group using \textit{EvoLLM} to verify specificity.}
\label{tab:main_robustness}
\end{table*}

\noindent\textbf{Evaluation Metrics.} We evaluate fingerprint performance using the Fingerprint Success Rate (FSR), defined as the fraction of queries that trigger the target response $y_{\text{target}}$. We report FSR on both the optimization trigger set and unseen model variants to assess in-distribution effectiveness and cross-model generalization, respectively.

\subsection{Main Results}
\label{sec:main-results}

\subsubsection{Robustness}
\label{sec:robustness}

\noindent\textbf{Effect of Multi-Model Joint Optimization.} 
Intuitively, using multiple homologous models for joint optimization can improve the robustness of the trigger set against downstream model modifications, including Supervised Fine-Tuning (SFT), Reinforcement Learning from Human Feedback (RLHF), model merging, and pruning. As shown in Tables \ref{tab:main_robustness}, \ref{tab:merging_robustness}, and \ref{tab:pruning_robustness}, compared to the fingerprint optimized solely on the base model (\texttt{Base (64)}), the \texttt{Base+Homologous Model}\footnote{The reported FSR for this configuration is comprehensively calculated from trigger sets optimized separately on combinations of Llama-2 and different individual homologous variants. For detailed, non-aggregated results, please refer to Figures \ref{fig:sft_rlhf_heatmap}, \ref{fig:merging_heatmap_full}, and \ref{fig:pruning_detailed}.} configuration demonstrates consistently higher FSR. For instance, it increases the FSR on the \textit{Vicuna} SFT variant from 0.14 to 0.31, on the \textit{DPO} alignment model from 0.34 to 0.46, and on the \textit{Taylor 5\%} pruned model from 0.14 to 0.36. This suggests that during the joint optimization process, the trigger perceives and anchors to the commonalities shared among multiple models rather than overfitting to the original base model, thereby enhancing its generalization capabilities.

\noindent\textbf{Effect of Multi-Template Joint Optimization.} 
Surprisingly, joint optimization using multiple chat templates---even when applied exclusively to a single base model---also significantly improves the trigger's robustness against downstream model modifications, including SFT, RLHF, merging, and pruning. The \texttt{Base+Chat Template}\footnote{Similarly, the reported FSR here is an aggregated result derived from trigger sets jointly optimized on the base model combined with different chat templates. The detailed numerical breakdown can also be found in Figures \ref{fig:sft_rlhf_heatmap}, \ref{fig:merging_heatmap_full}, and \ref{fig:pruning_detailed}.} configuration achieves notable FSR gains across various evaluation scenarios compared to the baseline. For example, it boosts the FSR on \textit{Task Arithmetic} merging from 0.58 to 0.82 and on the \textit{PPO-Pol} model from 0.80 to 0.91. We hypothesize that this phenomenon occurs because the multi-template optimization forces the trigger to capture the target model's intrinsic comprehension of the query. This fundamental comprehension capability is deeply embedded in the model and is consequently inherited by downstream variants, allowing the fingerprint to persist despite extensive parameter modifications.

\begin{table*}[ht]
\footnotesize
\centering
\setlength{\tabcolsep}{3.5pt} 
\renewcommand{\arraystretch}{1.1}
\resizebox{\textwidth}{!}{
    \begin{tabular}{l|c|cccccccc|c}
    \toprule
    \multirow{2}{*}{\textbf{Task Combination}} & \textbf{Anchor} & \multicolumn{8}{c|}{\textbf{Llama-2 + WizardMath Merges (8 Strategies)}} & \textbf{Existing Model} \\ \cmidrule(lr){2-2} \cmidrule(lr){3-10} \cmidrule(lr){11-11}
     & \textbf{Llama-2} & \textbf{Bread.} & \textbf{B-Ties} & \textbf{D-Task} & \textbf{D-Ties} & \textbf{Della} & \textbf{Del-Task} & \textbf{Task} & \textbf{Ties} & \textbf{FuseLLM} \\ \midrule
    Base (32) & 0.54 & 0.30 & 0.28 & 0.06 & 0.14 & 0.10 & 0.10 & 0.52 & 0.28 & 0.22 \\
    Base (64) & 0.78 & 0.42 & 0.46 & 0.24 & 0.26 & 0.18 & 0.18 & 0.58 & 0.38 & 0.50 \\
    Base+Homologous Model & 0.83 & 0.58 & 0.61 & 0.33 & 0.33 & 0.33 & 0.32 & 0.70 & 0.59 & 0.62 \\
    Base+Chat Template & 0.90 & 0.62 & 0.58 & 0.38 & 0.34 & 0.35 & 0.35 & 0.82 & 0.58 & 0.69 \\ \midrule
    \textbf{Base+S2.7-P+[D+Z]} & \textbf{0.98} & \textbf{0.80} & \textbf{0.80} & \textbf{0.52} & \textbf{0.62} & \textbf{0.54} & \textbf{0.60} & \textbf{0.94} & \textbf{0.80} & \textbf{0.88} \\ \midrule \midrule
    \textit{Base+Non-Homologous Model} & 0.02 & 0.00 & 0.00 & 0.00 & 0.02 & 0.00 & 0.00 & 0.02 & 0.00 & 0.00 \\ \bottomrule
    \end{tabular}
}
\caption{\textbf{Robustness evaluation against Model Merging.} The table reports the FSR on merged models. The first group comprises 8 distinct merging strategies applied to combine Llama-2-7B and WizardMath-7B. The second group contains \textit{FuseLLM}, an existing off-the-shelf merged model.}
\label{tab:merging_robustness}
\end{table*}

\noindent\textbf{Synergy of Pruned Variants and Multi-Template Optimization.} 
Building upon these findings, combining the base model with a structurally pruned variant alongside multi-template joint optimization yields the most resilient fingerprints. The multi-template optimization further increases the robustness ceiling. As demonstrated by the \texttt{Base+S2.7-P+[D+Z]} configuration, this synergistic approach elevates the FSR across all settings. It recovers substantial performance on heavily modified models, surging from 0.08 to \textbf{0.60} on \textit{ChineseLlama}, from 0.18 to \textbf{0.54} on the complex \textit{Della} sparse merge, and from 0.14 to \textbf{0.72} under \textit{Taylor 5\%} pruning. Even on highly specialized domains like \textit{CodeLlama}, it re-establishes a detectable signal (0.12), proving its effectiveness in uncovering deep-seated homologous signatures. Furthermore, the FSR on the off-the-shelf \textit{FuseLLM} improves substantially from 0.50 to \textbf{0.88}. Additional experiments on Qwen2.5 at both 7B and 14B scales and on Ministral-3-8B reproduce the overall benefit of joint model/template optimization, with the pruned+[D+Z] setting achieving the strongest overall robustness (Appendix~\ref{sec:cross_family_appendix}). For complete experimental results, including all intermediate baselines and detailed failure case analysis under drastic pruning, please refer to Appendices \ref{sec:sft_rlhf_heatmap}, \ref{sec:appendix_merging}, and \ref{sec:appendix_pruning}.

\begin{table}[t]
\centering
\setlength{\tabcolsep}{5pt}
\renewcommand{\arraystretch}{1.1}
\resizebox{\columnwidth}{!}{
\begin{tabular}{l|cc|cc}
\toprule
\multirow{2}{*}{\textbf{Task Combination}} & \multicolumn{2}{c|}{\textbf{Random Pruning}} & \multicolumn{2}{c}{\textbf{Taylor Pruning}} \\ \cmidrule(lr){2-3} \cmidrule(lr){4-5} 
 & \textbf{5\%} & \textbf{10\%} & \textbf{5\%} & \textbf{10\%} \\ \midrule
Base (32) & 0.34 & 0.06 & 0.24 & 0.02 \\
Base (64) & 0.44 & 0.20 & 0.14 & 0.04 \\
Base+Homologous Model & 0.56 & 0.22 & 0.36 & 0.10 \\
Base+Chat Template & 0.59 & 0.27 & 0.38 & 0.11 \\ \midrule
\textbf{Base+S2.7-P+[D+Z]} & \textbf{0.78} & \textbf{0.38} & \textbf{0.72} & \textbf{0.14} \\ \midrule \midrule
\textit{Base+Non-Homologous Model} & 0.00 & 0.00 & 0.00 & 0.00 \\ \bottomrule
\end{tabular}
}
\caption{\textbf{Robustness against Model Pruning.} We evaluate the FSR on models compressed via Random and Taylor pruning at different sparsity ratios.}
\label{tab:pruning_robustness}
\end{table}

\begin{table*}[ht]
\centering
\setlength{\tabcolsep}{3.5pt}
\renewcommand{\arraystretch}{1.1}
\resizebox{\textwidth}{!}{
    \begin{tabular}{l|ccccccccc|cc}
    \toprule
    \multirow{2}{*}{\textbf{Task Combination}} & \multicolumn{9}{c|}{\textbf{System Prompt Variations}} & \multicolumn{2}{c}{\textbf{Character Dropping}} \\ \cmidrule(lr){2-10} \cmidrule(lr){11-12}
     & \textbf{Fastchat} & \textbf{Llama2} & \textbf{OpenAI} & \textbf{Hiking} & \textbf{Tax} & \textbf{Json} & \textbf{Mktg.} & \textbf{Shake.} & \textbf{Cust.} & \textbf{~~-5\%} & \textbf{-10\%} \\ \midrule
    Base (32) & 0.16 & 0.10 & 0.26 & 0.14 & 0.08 & 0.10 & 0.10 & 0.14 & 0.10 & 0.12 & 0.00 \\
    Base (64) & 0.22 & 0.22 & 0.48 & 0.32 & 0.18 & 0.20 & 0.24 & 0.20 & 0.30 & 0.10 & 0.02 \\
    Base+Homologous Model & 0.39 & 0.30 & 0.56 & 0.34 & 0.22 & 0.33 & 0.37 & 0.30 & 0.35 & 0.13 & 0.02 \\
    Base+Chat Template & 0.54 & 0.47 & 0.72 & 0.53 & 0.26 & 0.48 & 0.52 & 0.38 & 0.45 & 0.09 & 0.02 \\ \midrule
    \textbf{Base+S2.7-P+[D+Z]} & \textbf{0.96} & \textbf{0.84} & \textbf{0.90} & \textbf{0.88} & \textbf{0.72} & \textbf{0.86} & \textbf{0.90} & \textbf{0.82} & \textbf{0.86} & \textbf{0.26} & 0.00 \\ \midrule \midrule
    \textit{Base+Non-Homologous Model} & 0.00 & 0.00 & 0.00 & 0.00 & 0.02 & 0.00 & 0.02 & 0.02 & 0.00 & 0.00 & 0.00 \\ \bottomrule
    \end{tabular}
}
\caption{\textbf{Robustness against Input Perturbations.} We evaluate the FSR under two types of input distortions: (1) \textit{System Prompt Variations}, where the adversarial input is wrapped in diverse conversational templates; and (2) \textit{Character Dropping}, where 5\% or 10\% of characters in the prompt are randomly discarded to simulate noise.}
\label{tab:input_perturbations}
\end{table*}

\begin{figure*}[t]
    \centering
    \includegraphics[width=\textwidth]{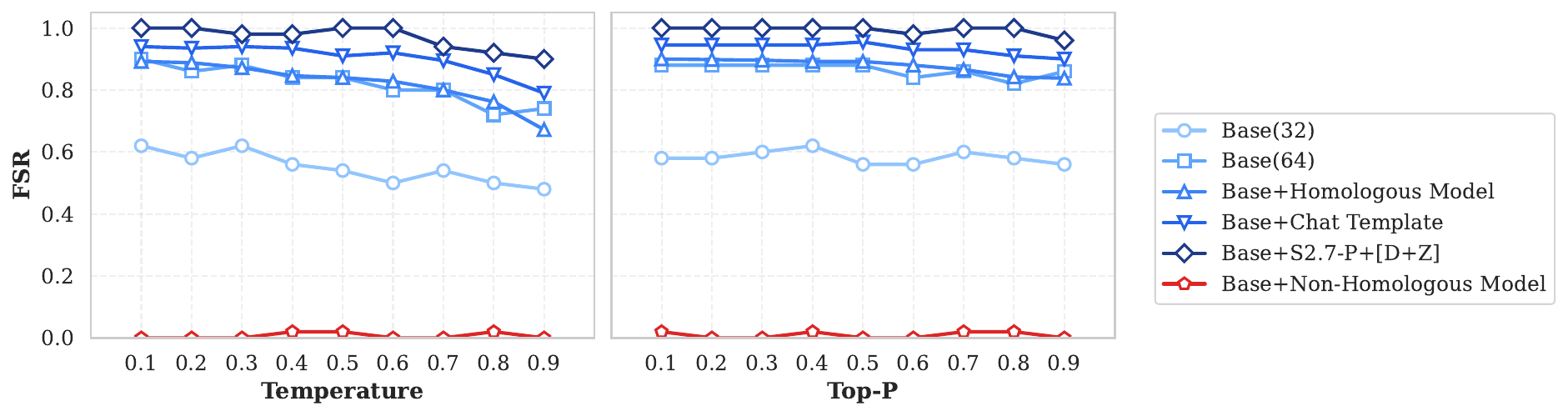}
    \caption{\textbf{Impact of generation hyperparameters on FSR.} The plots illustrate the stability of fingerprints under varying Temperature (left) and Top-P (right) settings. While baseline methods exhibit performance degradation at high temperatures (T>0.7) due to increased output entropy, our proposed \texttt{Base+S2.7-P+[D+Z]} strategy (green line) maintains consistently high success rates, demonstrating exceptional resilience to stochastic decoding noise.}
    \label{fig:hyperparams}
\end{figure*}

\noindent\textbf{Robustness to Input Perturbations.}
We evaluate fingerprint resilience against two common forms of real-world distortion: (1) system prompt variation, where the adversarial input is wrapped in diverse conversational contexts, and (2) transmission noise, where a fraction of characters is randomly dropped (Table \ref{tab:input_perturbations}). For system prompt evaluation, we adopt 9 diverse prompts—including general conversational agents and role-playing personas—following the TRAP framework \cite{gubri-etal-2024-trap} (see Appendix \ref{systemctl_prompt_appendix}). System prompts can severely disrupt adversarial suffixes; consequently, the baseline \texttt{Base (64)} performs poorly on complex wrappers like \textit{Tax} (0.18).

Intuitively, multi-template joint optimization (\texttt{Base+Chat Template}) improves the generalization of the trigger against system prompt modifications, as the trigger learns to operate across different conversational formats. Surprisingly, multi-model joint optimization (\texttt{Base+Homologous Model}) also significantly enhances robustness against prompt variations, indicating that exposing the trigger to diverse decision boundaries inherently increases its resilience to input phrasing modifications. 

Building upon these effects, our synergistic configuration (\texttt{Base+S2.7-P+[D+Z]}) demonstrates superior robustness. It achieves near-perfect FSRs on \textit{Fastchat} (\textbf{0.96}) and \textit{Marketing} (\textbf{0.90}), and maintains a high FSR of \textbf{0.72} even on the challenging \textit{Tax} prompt.
Regarding noise resilience, at a 5\% character drop rate, this configuration achieves an FSR of \textbf{0.26}, more than doubling the baseline (0.10) and indicating learned structural redundancy. However, performance collapses across all methods at a 10\% drop rate. Detailed breakdowns are provided in Appendix \ref{sec:appendix_perturbation}.

\noindent\textbf{Robustness to Generation Hyperparameters.} 
In real-world deployments, LLMs utilize stochastic decoding (e.g., varying Temperature and Top-P) for output diversity. We sweep Temperature ($T \in [0.1, 0.9]$) and Top-P ($p \in [0.1, 0.9]$) to evaluate stability (Figure \ref{fig:hyperparams}). While baseline methods degrade at high temperatures ($T > 0.7$) due to increased output entropy, our \texttt{Base+S2.7-P+[D+Z]} configuration maintains exceptional resilience, achieving an FSR $> 0.90$ even at $T=0.9$. Furthermore, performance remains virtually invariant across the Top-P spectrum, indicating our fingerprint successfully promotes the target response into the high-probability core of the distribution, rendering it immune to nucleus sampling truncation.

\subsubsection{Reliability}
\label{sec:reliability}

A reliable fingerprint must uniquely trigger the target model family while remaining inert on non-homologous models (i.e., minimizing False Positives). We evaluate the False Positive Rate (FPR) on 8 non-Llama-2 model families. The aggregated results are summarized in Table \ref{tab:reliability_fpr}.

The baseline (\texttt{Base(64)}) exhibits exceptional specificity, with an average FPR near zero (0.003). This confirms that single-model optimization exploits highly specific artifacts that do not transfer to disjoint architectures.
A key concern with our robustness-enhancing method is the potential trade-off with specificity. Our results indicate that while joint optimization slightly increases the FPR, it remains negligible. Our method (\texttt{Base+S2.7-P+[D+Z]}) maintains a low average FPR of 0.035 and a maximum of 0.08 (on \textit{Qwen2.5-7B}), confirming it rarely misidentifies non-homologous models. Detailed breakdowns are in Appendix \ref{sec:appendix_reliability}.

\begin{table}[t]
\footnotesize
\centering
\setlength{\tabcolsep}{5pt}
\renewcommand{\arraystretch}{1.1}
\resizebox{\columnwidth}{!}{
\begin{tabular}{l|cc}
\toprule
\textbf{Task Combination} & \textbf{Avg. FPR} & \textbf{Max. FPR} \\ \midrule
Base (32) & 0.005 & 0.02 \\
Base (64) & 0.003 & 0.02 \\
Base+Homologous Model & 0.008 & 0.02 \\
Base+Pruned Model & 0.030 & 0.04 \\
Base+Chat Template & 0.011 & 0.03 \\ \midrule
\textbf{Base+S2.7-P+[D+Z]} & \textbf{0.035} & \textbf{0.08} \\ \midrule \midrule
\textit{Base+Non-Homologous Model} & 0.000 & 0.00 \\ \bottomrule
\end{tabular}
}
\caption{\textbf{Reliability Evaluation.} Due to space constraints, we report the Average and Maximum FPR across 8 non-homologous models.}
\label{tab:reliability_fpr}
\end{table}

\begin{figure}[t]
    \centering
    \includegraphics[width=0.985\columnwidth]{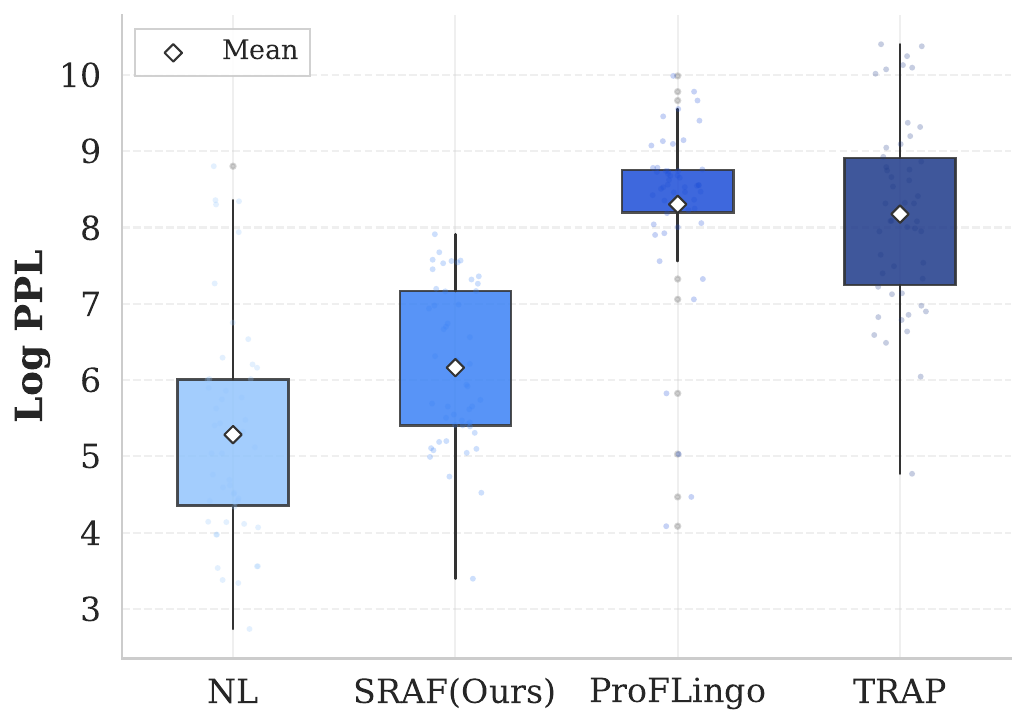} 
    \caption{\textbf{Log PPL Distribution.} \texttt{NL} represents original Natural Language prompts.}
    \label{fig:ppl_comparison}
\end{figure}

\subsubsection{Stealthiness}
\label{sec:exp_stealthiness}

Stealthiness is crucial for model fingerprinting because if a trigger is easily identifiable as an anomalous query, adversaries can simply filter it out to evade ownership verification. Perplexity (PPL) is a standard metric widely used to evaluate the stealthiness of such inputs \cite{xuCTCCRobustStealthy2025,xuEverTracerHuntingStolen2025}. A lower perplexity indicates that the token sequence closely follows the statistical patterns of natural language, making it harder for perplexity-based defense mechanisms to detect.

To demonstrate the effectiveness of our Perplexity Hiding technique, we measure the Log PPL using the GPT-2 tokenizer. We compare the trigger sets optimized by our method against existing baselines (TRAP \cite{gubri-etal-2024-trap} and ProFLingo \cite{10735575}), the standard \texttt{Base (64)} optimization without joint optimization, and the original Natural Language (NL) prompts. 

As shown in Figure \ref{fig:ppl_comparison}, existing methods (TRAP and ProFLingo) and the \texttt{Base (64)} without joint optimization exhibit significantly elevated perplexity levels, with Log PPL values clustering in the anomalous high range of 7 to 10. In contrast, SRAF with Perplexity Hiding demonstrates a substantial reduction in perplexity. Its distribution (the red cluster) largely overlaps with the Natural Language baseline. This confirms that embedding the adversarial tokens within a Markdown table structure effectively lowers PPL, making the trigger more aligned with natural language in terms of the model's statistical comprehension. A controlled sensitivity analysis further shows that table shape and header wording primarily affect PPL, whereas FSR remains stable when the total optimization budget is fixed at 64 tokens. Across the evaluated designs, average FSR varies by only 0.04, while balanced layouts and semantic headers achieve lower PPL; full results are reported in Appendix~\ref{sec:table_design_sensitivity}.

\section{Discussion}
\label{sec:discussion}

\begin{figure}[t]
    \centering
    \includegraphics[width=\columnwidth]{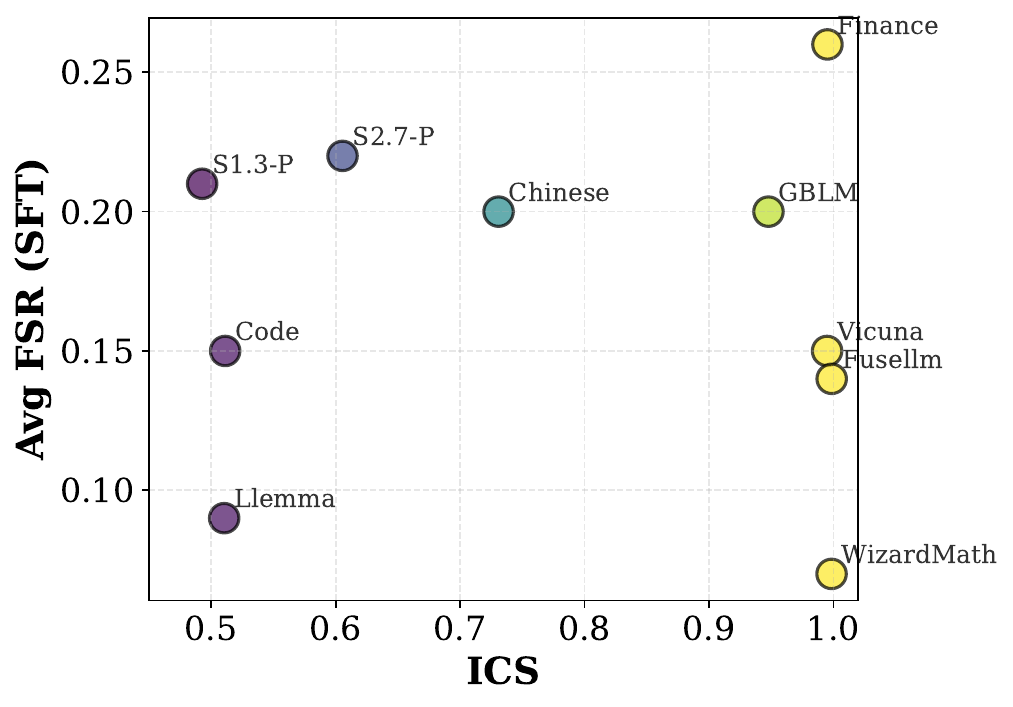}
    \caption{\textbf{Correlation between Model Similarity and Robustness on SFT tasks.} Each point represents a specific model variant used as a co-optimization target.}
    \label{fig:ics_sft}
\end{figure}

A common intuition in adversarial transferability is that fingerprints should transfer most effectively between models that share high parameter similarity. To investigate this, we analyze the relationship between Invariant Terms’ Cosine Similarity (ICS) \cite{zeng2025hurefhumanreadablefingerprintlarge} —a weight-based similarity metric—measured relative to Llama-2-7B, and the FSR achieved when using that model as a co-optimization target. Figure \ref{fig:ics_sft} illustrates this relationship on SFT variants. The results reveal a counter-intuitive phenomenon: \textbf{parameter space similarity is a poor predictor of adversarial robustness.}

As shown in the plot, \textit{WizardMath-7B} exhibits an ICS near 1.0, indicating its weights are numerically very close to the base model. However, the fingerprint optimized jointly with it yields a remarkably low FSR ($<0.10$). Conversely, the pruned model S2.7-P has a low ICS ($\approx 0.6$) due to significant structural compression. Yet, it achieves a high average FSR ($\approx 0.22$). This observation justifies our selection of structurally distinct models for joint optimization. This conclusion also holds for RLHF and Merged model variants, as demonstrated in Appendix \ref{sec:appendix_ics} (Figure \ref{fig:appendix_scatters}).

\section{Conclusion}
In this work, we presented \textbf{SRAF}, a robust and stealthy adversarial fingerprinting framework for the copyright protection of LLMs. Addressing the critical shortcomings of existing black-box methods, SRAF introduces two key technical innovations: a multi-task optimization objective that generalizes across model families and prompt templates, and a perplexity-hiding mechanism leveraging Markdown table structures.
Our extensive evaluation demonstrates that SRAF maintains high identification success rates even when the target model undergoes post-processing scenarios where baseline methods typically fail. Furthermore, SRAF exhibits exceptional stealthiness, generating fingerprints that are statistically indistinguishable from natural language, thereby resisting perplexity-based filtering. SRAF establishes a new standard for practical, non-invasive model ownership verification in black-box environments.

\section*{Limitations}

Despite its strong empirical performance, SRAF has several limitations. First, although our evaluation covers three open-weight model families
and parameter scales from 7B to 14B, the most comprehensive modification suite remains centered on Llama-2-7B. Substantially larger models,
additional architectures, and closed-source commercial APIs remain unexplored.

Second, the chat templates used in multi-template optimization are only representative rather than exhaustive. Real-world deployments may adopt customized system prompts, proprietary formatting rules, or dynamically changing wrappers. Therefore, although SRAF improves robustness against several unseen prompt variations, it cannot guarantee invariance to arbitrary templates. Third, while joint optimization with structurally pruned models empirically achieves the best robustness, the reason behind this phenomenon lacks a formal theoretical explanation. We hypothesize that pruning encourages the fingerprint to rely on more invariant model-family features, but this remains to be rigorously verified.

Finally, SRAF is still limited under extreme structural modifications. When a model undergoes aggressive L1/L2-norm structured pruning or substantial architectural changes, the shared homologous features between the anchor and derivative models may be disrupted, causing the adversarial fingerprint to lose effectiveness.

\section*{Acknowledgments}

The authors would like to acknowledge the support of the National Natural Science Foundation of China under Grant No. 62376246.

\bibliography{custom}

\appendix

\section{Expirement Results}
\subsection{Detailed Analysis of Robustness to SFT and RLHF}
\label{sec:sft_rlhf_heatmap}

\begin{figure*}[ht]
    \centering
    \includegraphics[width=\textwidth]{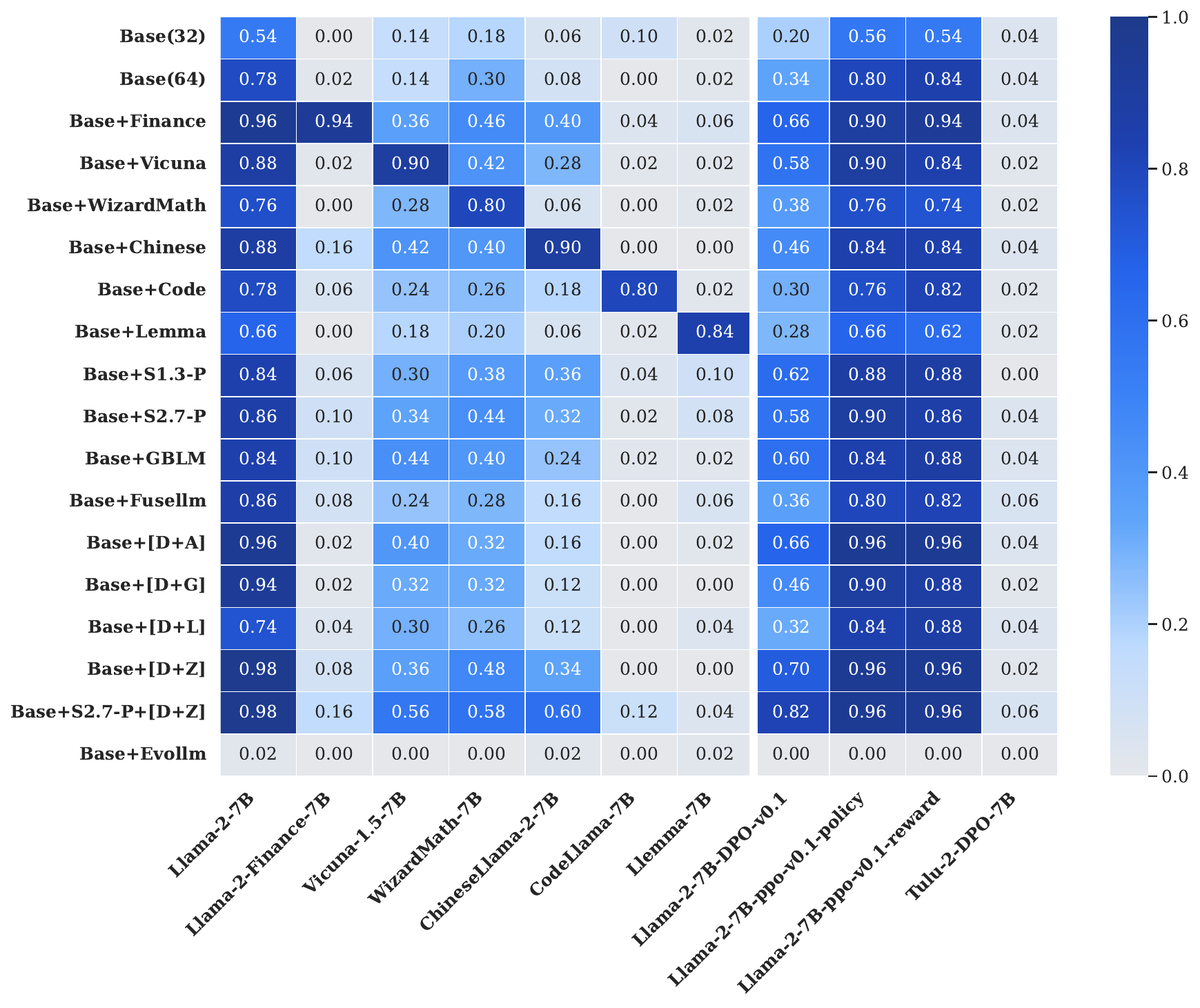} 
    \caption{\textbf{Complete Robustness Heatmap to SFT and RLHF.} The figure displays the FSR for all evaluated optimization strategies (rows) against the full suite of target models (columns). While single-target joint optimization (e.g., \texttt{Base+Finance}) solves specific domain transfers, it lacks cross-domain universality. Our proposed method, \texttt{Base+S2.7-P+[D+Z]}, achieves the best global robustness across SFT and RLHF variants. The final row, \texttt{Base+EvolIm}, acts as a control, showing zero success on non-homologous models.}
    \label{fig:sft_rlhf_heatmap}
\end{figure*}

\begin{figure*}[ht]
    \centering
    \includegraphics[width=\textwidth]{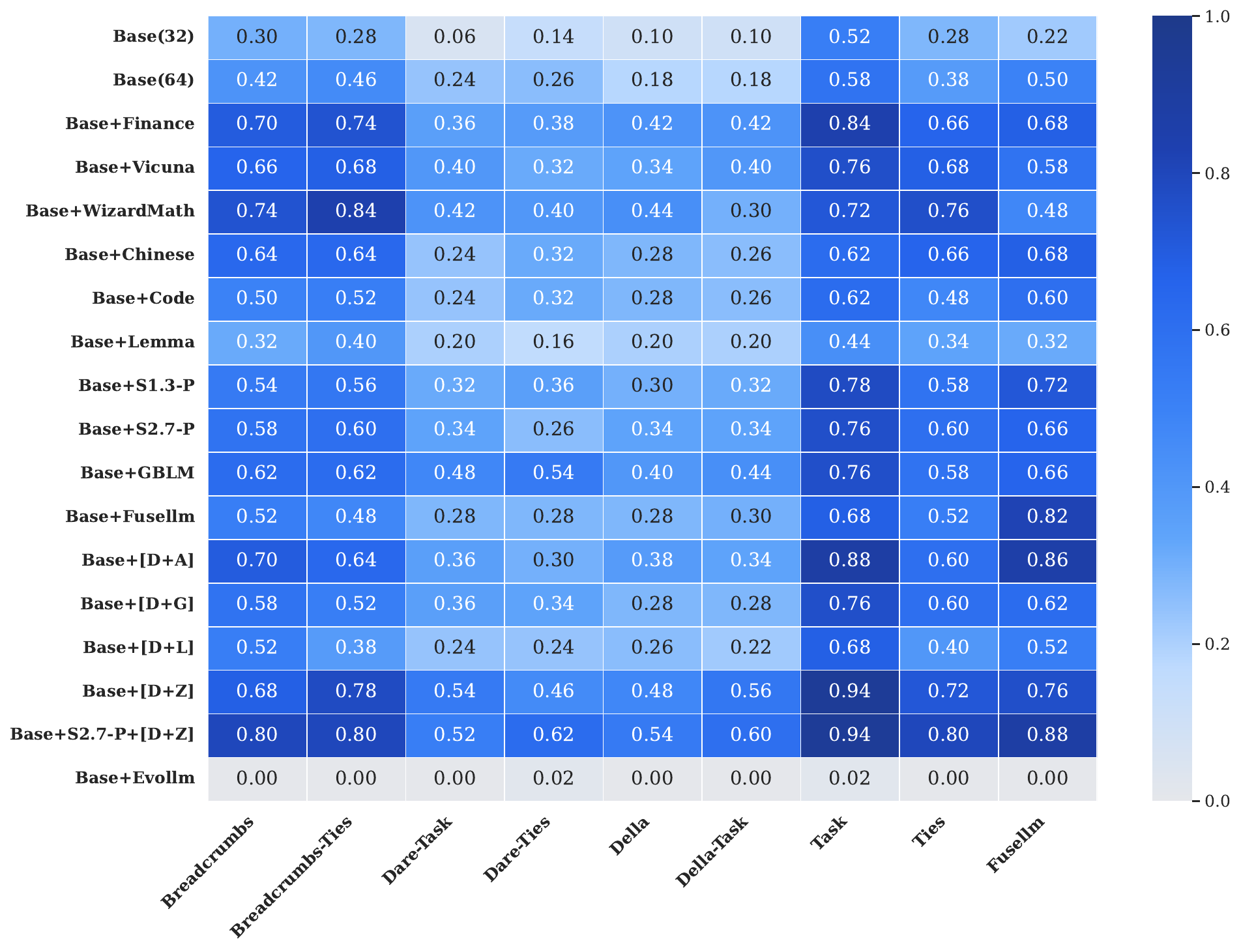} 
    \caption{\textbf{Complete Heatmap of Robustness against Model Merging.} The figure details the transferability of fingerprints optimized under various settings (rows) to merged models (columns). While optimizing on the specific constituent model (\texttt{Base+WizardMath}) works well for this specific merge pair, our generalized approach \texttt{Base+S2.7-P+[D+Z]} achieves the highest global robustness, particularly on challenging sparse merges like \textit{Della} and \textit{Dare}.}
    \label{fig:merging_heatmap_full}
\end{figure*}

In Section \ref{sec:robustness}, we presented a summarized evaluation of fingerprint robustness across SFT and RLHF variants. Here, we provide the comprehensive experimental results in Figure \ref{fig:sft_rlhf_heatmap}, illustrating the detailed performance of distinct optimization configurations. The full heatmap reveals several granular insights that motivate our final choice of the \texttt{Base+S2.7-P+[D+Z]} configuration:

\begin{itemize}[leftmargin=*, nolistsep]
    \item \textbf{Impact of Fingerprint Length:} Comparing \texttt{Base(32)} and \texttt{Base(64)}, we observe that increasing the adversarial string length from 32 to 64 tokens universally improves the FSR on the anchor model and easy targets (e.g., PPO variants). However, increased length alone is insufficient to overcome the domain shift in SFT models like \textit{CodeLlama} or \textit{Llama-2-Finance}.
    
    \item \textbf{Specificity of Single-Target Optimization:} Rows such as \texttt{Base+Finance} and \texttt{Base+Code} demonstrate the ``catastrophic forgetting'' phenomenon in adversarial transferability. While optimizing jointly with the Finance model yields a high FSR on \textit{Llama-2-Finance-7B} (0.94), it fails to generalize to other domains like \textit{CodeLlama} (FSR 0.04). Similarly, \texttt{Base+Code} achieves 0.80 on the code variant but performs poorly on finance models. This highlights that incorporating a single SFT variant into the optimization loop creates a narrow robustness profile.
    
    \item \textbf{Generalization via Pruning and Templates:} The configuration \texttt{Base+S2.7-P+[D+Z]} stands out by maintaining high FSR across disjoint domains (Finance, Code, Math) and distinct alignment algorithms (DPO, PPO). This confirms that optimizing against a structurally pruned model (\textit{Sheared-Llama-2.7B}) combined with diverse template formats prevents overfitting to specific model weights, forcing the algorithm to find a truly universal ``family signature''.
\end{itemize}

\subsection{Detailed Analysis of Robustness against Model Merging}
\label{sec:appendix_merging}

Figure \ref{fig:merging_heatmap_full} presents the comprehensive evaluation of fingerprint robustness across various model merging scenarios. The target models (columns) are constructed by merging Llama-2-7B with WizardMath-7B using 8 different strategies, alongside the off-the-shelf FuseLLM-7B.

Beyond the main findings discussed in Section \ref{sec:robustness}, the full heatmap provides additional insights into the interaction between optimization sources and merging mechanisms:

\begin{itemize}[leftmargin=*, nolistsep]
    \item \textbf{Effectiveness of Constituent Model Optimization:} We observe that \texttt{Base+WizardMath} performs surprisingly well, achieving an FSR of 0.84 on \textit{Breadcrumbs-Ties}. This is expected because the target models are explicitly composed of Llama-2 and WizardMath weights. Optimizing on one of the constituent models (WizardMath) naturally yields a fingerprint that transfers well to the child merge. However, this assumes knowledge of the merge components, whereas \texttt{Base+S2.7-P+[D+Z]} achieves superior results.
    
    \item \textbf{Pruning Mimics Sparsification:} Strategies involving \texttt{Base+S2.7-P} (rows 11 and 16) show a distinct advantage on ``sparse'' merging targets like \textit{Dare-Task} and \textit{Della}. Since Dare and Della techniques employ delta-parameter sparsification, optimizing against a pruned model (Sheared-Llama) likely forces the fingerprint to rely on the most salient, high-magnitude weights that survive both pruning and merging sparsification processes.
    
    \item \textbf{Template Generalization:} The \texttt{Base+[D+Z]} configuration (row 15) proves to be a strong baseline, particularly on \textit{Task Arithmetic} (0.94). This suggests that enforcing robustness across Default and Zero-shot templates helps the fingerprint bypass the specific instruction-tuning artifacts that might be disrupted during the merging process.
\end{itemize}

\subsection{Extended Pruning Robustness Analysis}
\label{sec:appendix_pruning}

In Section \ref{sec:robustness}, we summarized the robustness of SRAF against Random and Taylor pruning. Here, we expand this analysis to cover a wider range of optimization strategies and investigate the limits of fingerprint transferability under aggressive compression methods, including L1-norm, L2-norm, and specialized compression frameworks (Wanda, GBLM, Sheared).

\begin{figure}[t]
    \centering
    \includegraphics[width=\columnwidth]{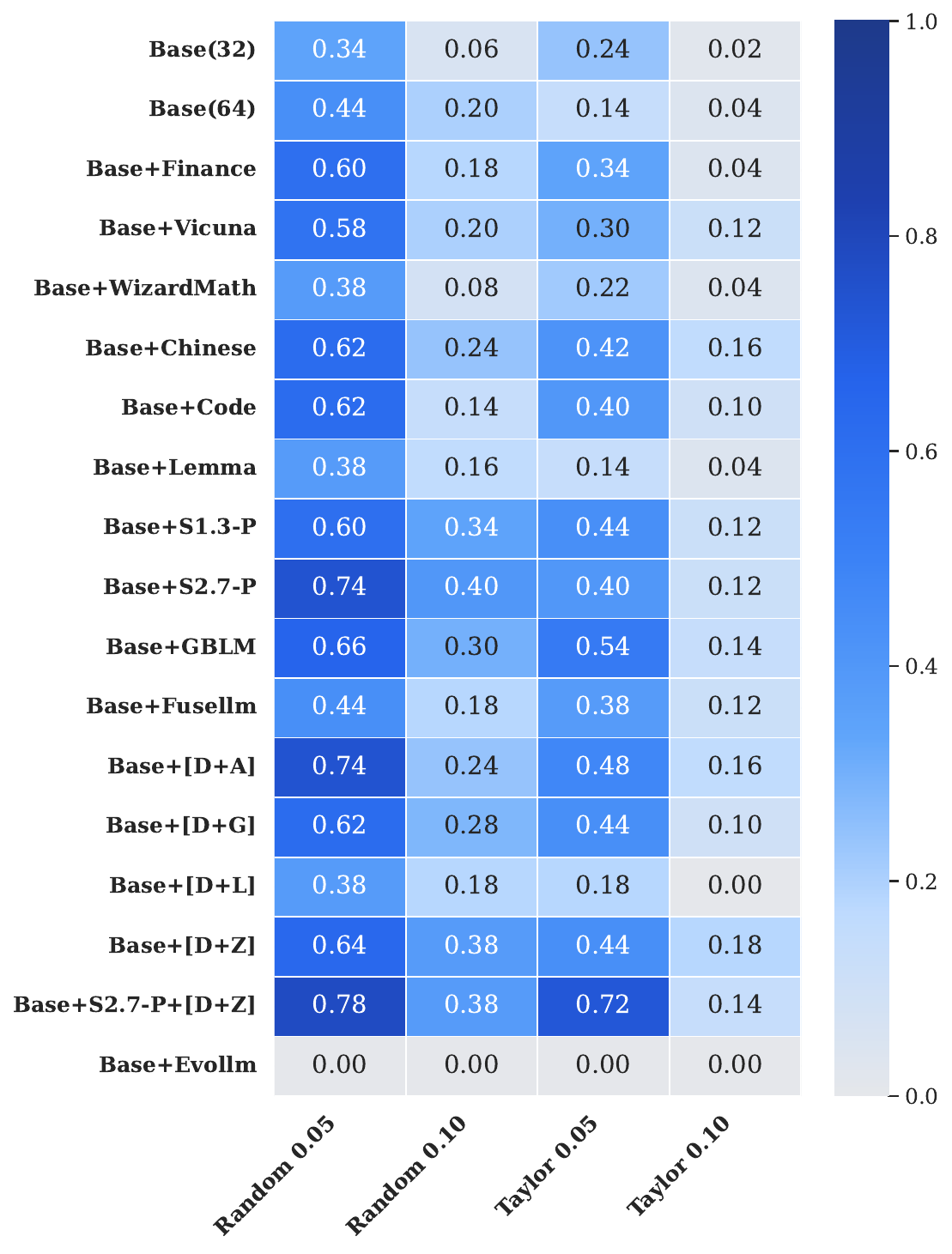} 
    \caption{\textbf{Detailed Robustness against Standard Pruning.} The heatmap shows the FSR for various optimization strategies against Random and Taylor pruning. \texttt{Base+S2.7-P+[D+Z]} consistently outperforms other methods, demonstrating that joint optimization with a pruned model enhances resilience against importance-based parameter removal.}
    \label{fig:pruning_detailed}
\end{figure}

\noindent\textbf{Performance on Standard Pruning.}
Figure \ref{fig:pruning_detailed} details the FSR across various optimization settings for Random and Taylor pruning. 
The results highlight the efficacy of incorporating pruned variants during training. For example, while the baseline \texttt{Base(64)} struggles with Taylor 5\% pruning (FSR 0.14), the specialized configuration \texttt{Base+S2.7-P} (trained with Sheared-Llama-2.7B) recovers the performance to 0.40. Notably, our final joint strategy \texttt{Base+S2.7-P+[D+Z]} achieves the highest robustness, reaching 0.72 on Taylor 5\% and 0.78 on Random 5\%, confirming that template diversity aids in finding robust features that survive compression.

\begin{figure*}[ht]
    \centering
    \includegraphics[width=\textwidth]{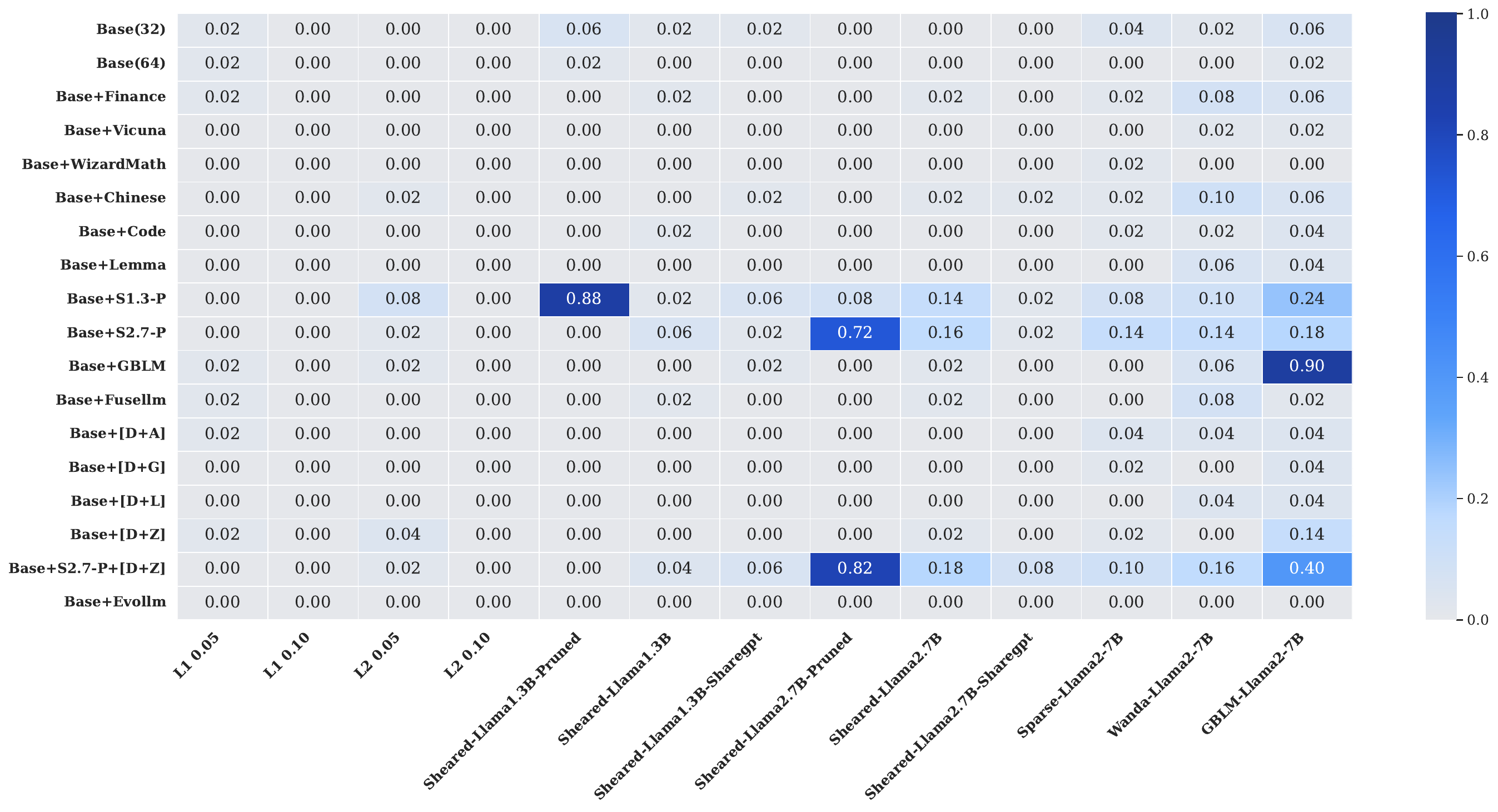} 
    \caption{\textbf{Robustness Limits under Aggressive Compression.} The heatmap evaluates FSR against L1/L2 pruning and specific compressed models (Sheared, Wanda, GBLM). Most cells are near zero, indicating that zero-shot transfer fails when the architecture is drastically altered. High success rates are only observed when the specific compressed model is included in the optimization (e.g., \texttt{Base+GBLM} attacking \textit{GBLM-Llama2-7b}), highlighting the specificity required for these targets.}
    \label{fig:pruning_failure}
\end{figure*}

\noindent\textbf{Failure Cases under Aggressive Compression.}
Figure \ref{fig:pruning_failure} illustrates the performance on ``hard'' pruning targets, including L1/L2 structured pruning and specific compressed model families. Two key observations emerge:
\begin{enumerate}[leftmargin=*, nolistsep]
    \item \textbf{Failure of Zero-Shot Transfer:} The columns for L1 and L2 pruning are almost entirely zero (FSR $\approx$ 0.00) across all optimization methods. This indicates that norm-based structured pruning alters the model's decision boundary so fundamentally that the adversarial trajectory learned on the dense model becomes invalid. Similarly, transfer to unseen architectures like \textit{Wanda-Llama2-7B} or \textit{Sparse-Llama2-7B} is negligible.
    
    \item \textbf{Specificity of Compressed Fingerprints:} High FSR values appear only on the diagonal where the optimization target matches the evaluation model. for instance, \texttt{Base+S1.3-P} achieves 0.88 on \textit{Sheared-Llama1.3B}, and \texttt{Base+GBLM} achieves 0.90 on \textit{GBLM-Llama2-7B}. However, these fingerprints do not transfer to other compressed variants. This suggests that aggressive compression creates distinct ``model islands'' with unique vulnerabilities that do not overlap with the original model or other compressed variants.
\end{enumerate}

\subsection{Detailed Evaluation of Input Perturbation Robustness}
\label{sec:appendix_perturbation}

\begin{figure*}[h]
    \centering
    \includegraphics[width=\textwidth]{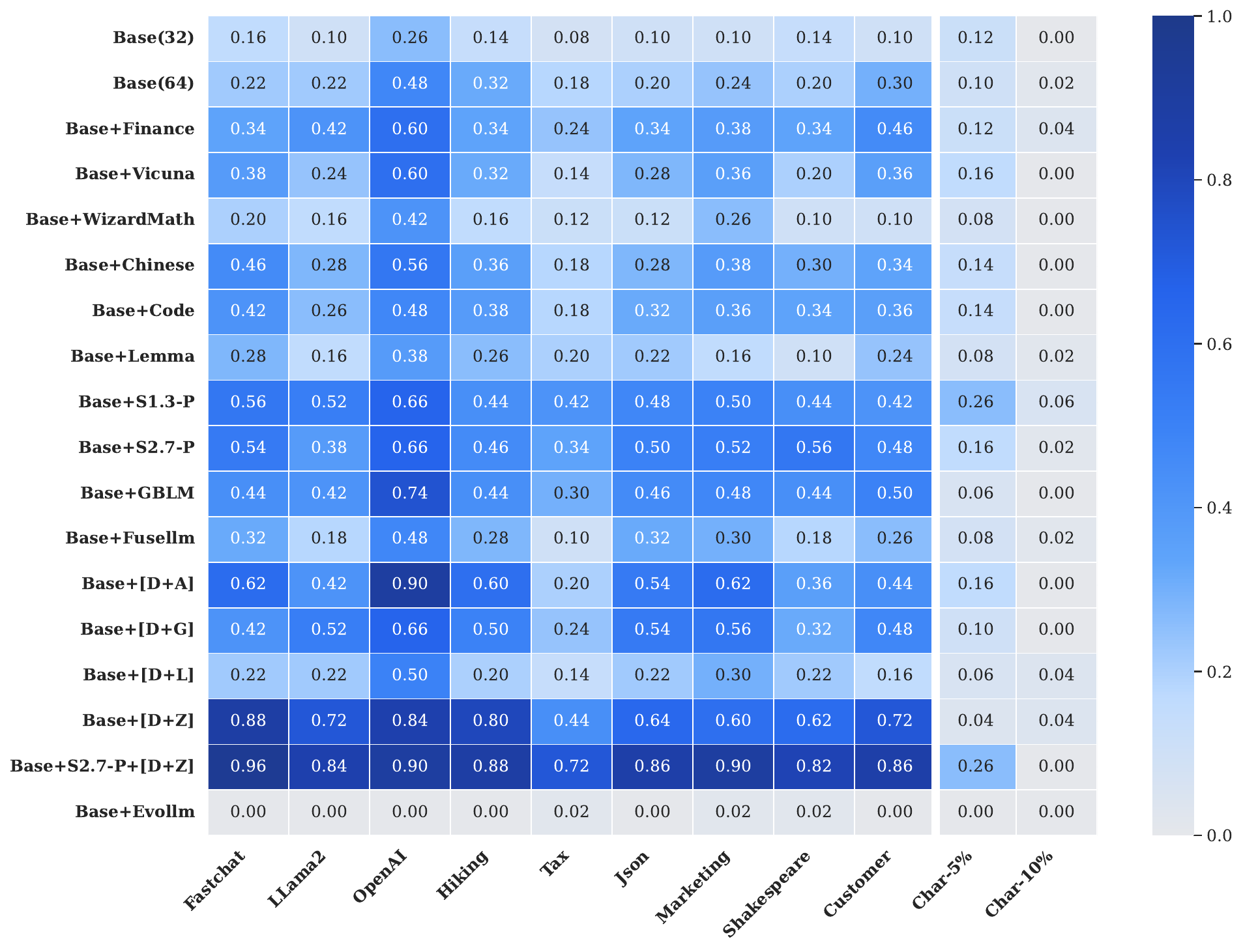}
    \caption{\textbf{Complete Robustness Heatmap against Input Perturbations.} The figure details the FSR of all optimization variants against 9 diverse system prompts and 2 noise levels. While template-based optimization (e.g., \texttt{Base+[D+Z]}) improves resilience to system prompts, it fails under character dropping. Our joint method (\texttt{Base+S2.7-P+[D+Z]}) provides the most comprehensive protection, maintaining high FSRs across complex wrappers and moderate noise levels.}
    \label{fig:perturbation_heatmap}
\end{figure*}

In Section \ref{sec:robustness}, we summarized the robustness of our method against system prompts and character dropping. Figure \ref{fig:perturbation_heatmap} presents the complete experimental results, offering a granular view of how each optimization strategy handles input distortions. Several key observations can be drawn from the full heatmap:

\begin{itemize}[leftmargin=*, nolistsep]
    \item \textbf{Ineffectiveness of Domain Adaptation:} Strategies based solely on SFT models (e.g., \texttt{Base+Finance}, \texttt{Base+Code}) show limited improvement over the baseline when facing system prompt variations. For example, \texttt{Base+Code} achieves only 0.42 on \textit{Fastchat} and 0.18 on \textit{Tax}. This indicates that simply training on domain-shifted weights does not inherently provide robustness against prompt wrapping.
    
    \item \textbf{Impact of Chat Templates:} As expected, strategies explicitly optimizing for chat templates (e.g., \texttt{Base+[D+Z]}) perform significantly better on system prompts, which share structural similarities with chat formats. \texttt{Base+[D+Z]} achieves a high FSR of 0.88 on \textit{Fastchat}. However, it still exhibits instability on difficult prompts like \textit{Tax} (0.44) and suffers under noise conditions (0.04 on Char-5\%).
    
    \item \textbf{Synergy of Joint Optimization:} The optimal configuration, \texttt{Base+S2.7-P+[D+Z]}, combines the structural invariance learned from the pruned model with the format robustness of chat templates. This synergy yields the highest success rates across the board (e.g., 0.72 on \textit{Tax}, 0.26 on Char-5\%), confirming that combining parameter-space constraints (pruning) with input-space diversity (templates) is essential for maximum robustness.
\end{itemize}

\subsection{Detailed Reliability Analysis}
\label{sec:appendix_reliability}

\begin{figure*}[h]
    \centering
    \includegraphics[width=\textwidth]{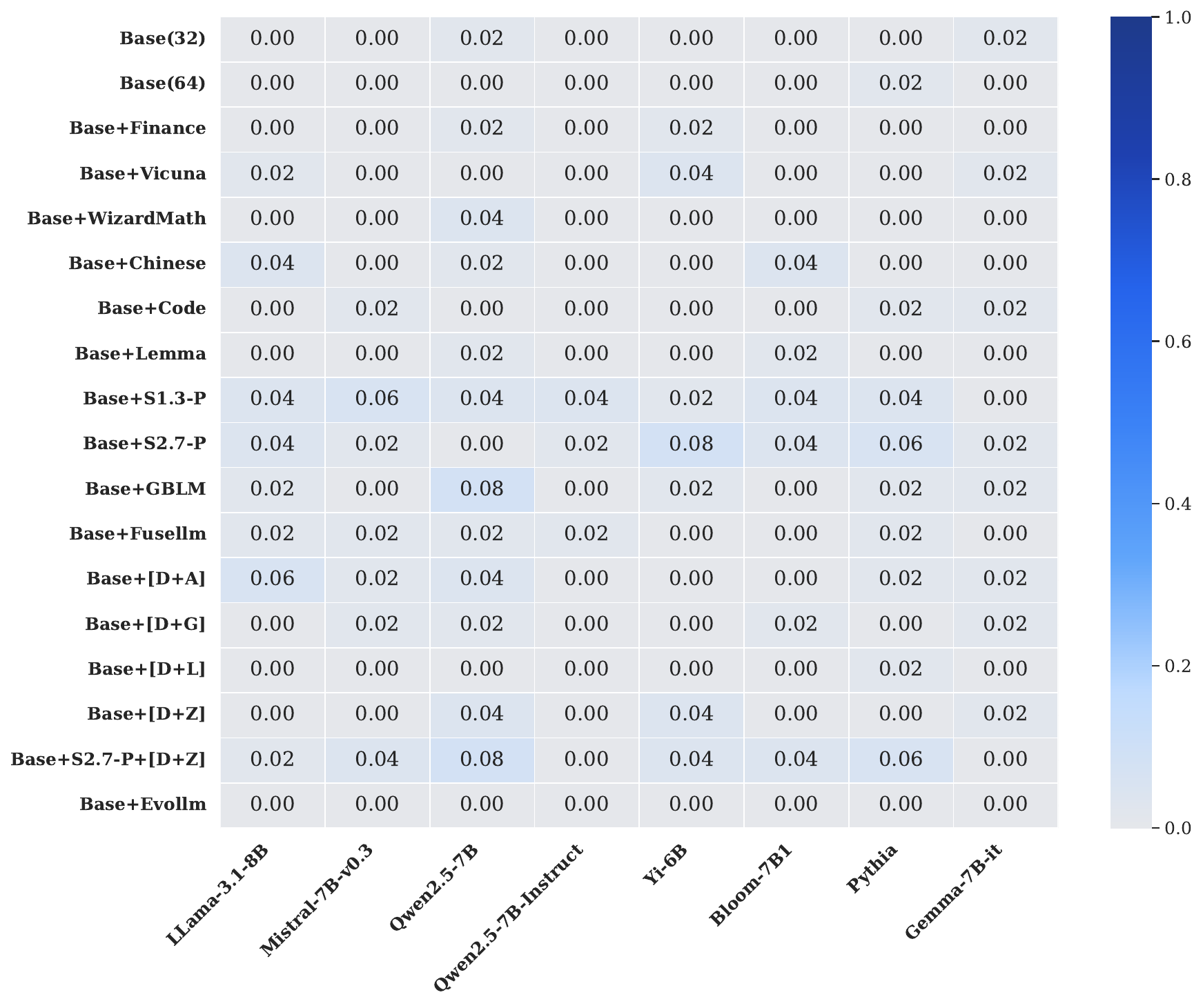} 
    \caption{\textbf{Complete Reliability Heatmap (False Positive Analysis).} The figure details the FSR on non-homologous models. The values represent the False Positive Rate (FPR). The predominance of near-zero values (grey/light blue) confirms high specificity.}
    \label{fig:reliability_heatmap}
\end{figure*}

Complementing the summary provided in Section \ref{sec:reliability}, we present the complete false positive evaluation in Figure \ref{fig:reliability_heatmap}. This heatmap visualizes the FSR across 8 non-target model families for all optimization configurations. The comprehensive data supports three key conclusions regarding the specificity of our method:

\begin{itemize}[leftmargin=*, nolistsep]
    \item \textbf{Universal Specificity:} Across nearly all cells, the FSR remains strictly below 0.05. Even for models with high capabilities, the adversarial fingerprints fail to trigger the target response. This confirms that the optimized suffixes rely on specific weight artifacts of the Llama-2 family rather than generic semantic loopholes or ``universal jailbreak'' patterns.
    
    \item \textbf{Differentiation from Llama-3:} A critical test case is \textit{Llama-3.1-8B}, which shares the same tokenizer and architectural lineage as Llama-2 but possesses different weights. The heatmap shows negligible activation (max 0.06) on Llama-3.1. This effectively demonstrates that our fingerprinting method is \textit{version-specific}, allowing for precise distinction between model generations within the same provider.
    
    \item \textbf{Impact of Aggressive Optimization:} Strategies involving pruned models or template generalization exhibit a marginal increase in FPR compared to single-domain adaptation. However, this increase is statistically insignificant compared to the operational detection threshold, ensuring that the gain in robustness does not come at the cost of reliability.
\end{itemize}

\subsection{Extended Correlation Analysis on RLHF and Merging}
\label{sec:appendix_ics}

\begin{figure*}[t]
    \centering
    \begin{subfigure}{0.48\textwidth}
        \centering
        \includegraphics[width=\linewidth]{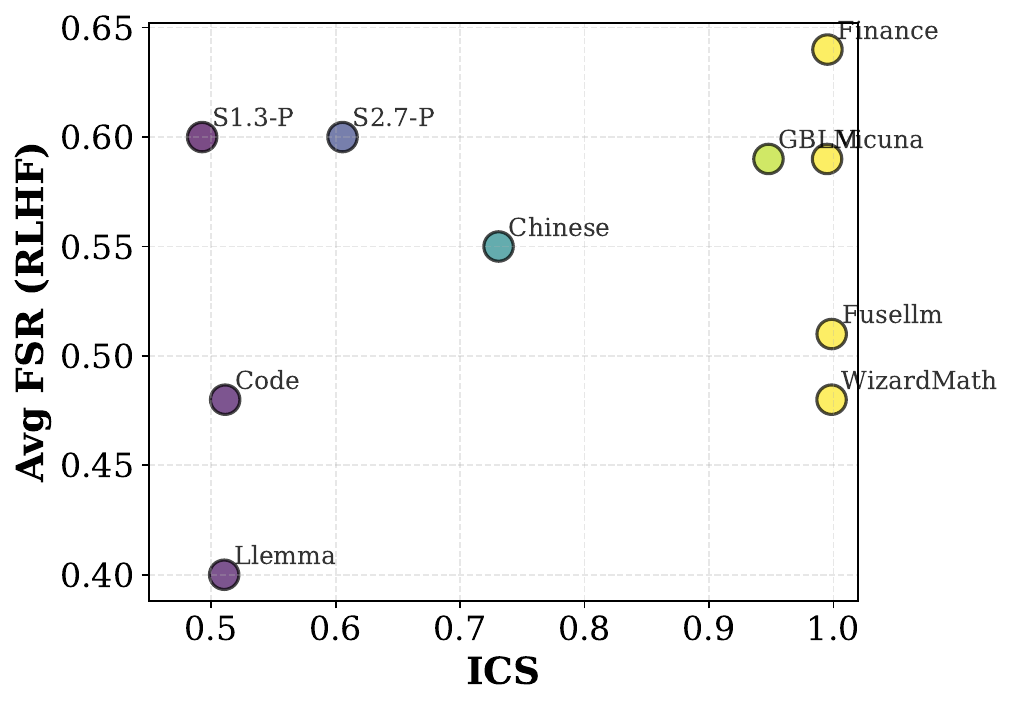} 
        \caption{Robustness on RLHF Variants}
        \label{fig:scatter_rlhf}
    \end{subfigure}
    \hfill
    \begin{subfigure}{0.48\textwidth}
        \centering
        \includegraphics[width=\linewidth]{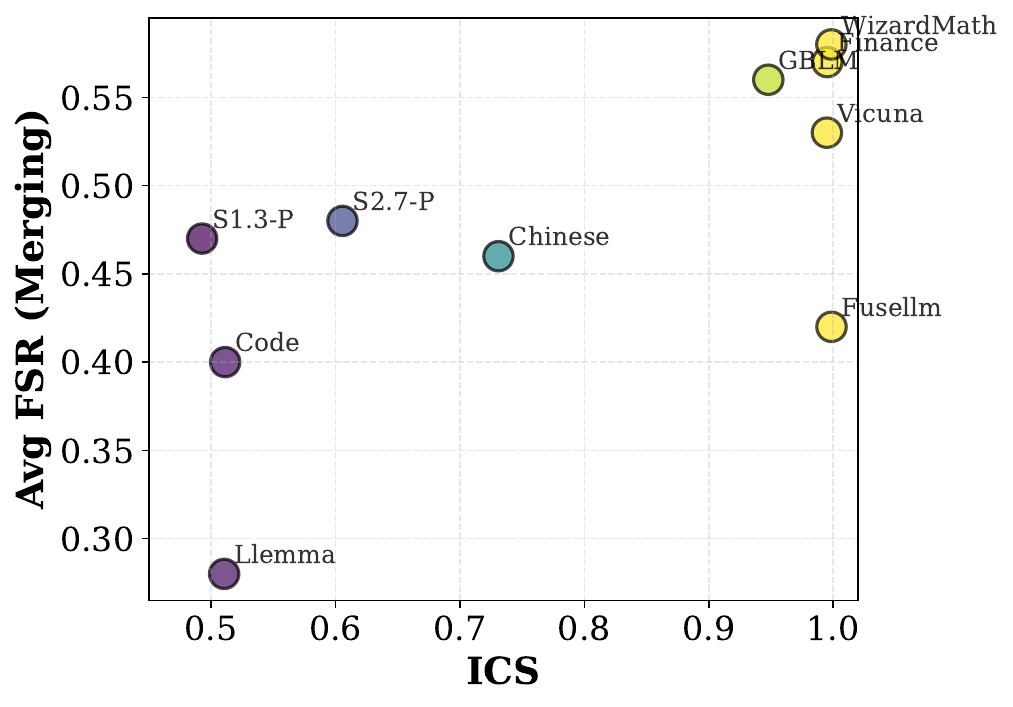} 
        \caption{Robustness on Merged Variants}
        \label{fig:scatter_merging}
    \end{subfigure}
    
    \caption{\textbf{Extended Analysis of ICS vs. FSR.} The scatter plots for RLHF (a) and Merging (b) tasks confirm that high parameter similarity (ICS) is not a prerequisite for effective joint optimization.}
    \label{fig:appendix_scatters}
\end{figure*}

In Section \ref{sec:discussion}, we demonstrated the lack of strong correlation between parameter similarity (ICS) and robustness on SFT models. Here, we extend this analysis to RLHF and Merged model variants to verify the generalizability of this finding. Figure \ref{fig:appendix_scatters} presents the ICS-FSR relationship for RLHF target and Merging targets. The trends align consistently with our previous observations:

\begin{itemize}[leftmargin=*, nolistsep]
    \item \textbf{RLHF Consistency:} In the RLHF setting, \textit{S2.7-P} continues to outperform models with higher parameter similarity, such as \textit{ChineseLlama} and \textit{CodeLlama}. The outlier status of \textit{WizardMath} (high ICS, low FSR) persists, confirming that global parameter proximity does not guarantee the preservation of alignment-invariant vulnerabilities.
    \item \textbf{Merging Consistency:} For merged models, while \textit{Finance} shows strong performance, the pruned variants (\textit{S2.7-P}, \textit{S1.3-P}) maintain competitive robustness despite their low ICS scores.
\end{itemize}

\subsection{Cross-Family and Cross-Scale Validation}
\label{sec:cross_family_appendix}

We conduct additional experiments to determine whether the robustness improvements observed on Llama-2 generalize across model families and
parameter scales. All experiments use the default $2\times4$ Markdown table and a 64-token optimization budget. Importantly, fingerprints are
optimized separately for each family and evaluated only on the anchor and post-trained variants from that family; we do not assume that a single
fingerprint transfers across unrelated families.

For Qwen2.5-7B, we evaluate transfer to an instruction-tuned variant, a merged Math-TIES variant, and a Wanda-pruned variant. For Qwen2.5-14B and
Ministral-3-8B, we compare five common optimization settings: base-only, multi-template [D+Z], joint optimization with an instruction-tuned model, joint optimization with a pruned model, and the combination of the pruned model and [D+Z].

\begin{table*}[t]
\centering
\footnotesize
\renewcommand{\arraystretch}{1.05}

\begin{tabular*}{0.86\textwidth}
{@{\extracolsep{\fill}}l|cccc@{}}

\multicolumn{5}{c}{\textit{(a) Qwen2.5-7B}} \\
\toprule
\textbf{Optimization Setting}
& \textbf{Base}
& \textbf{Instruct}
& \textbf{Math-TIES}
& \textbf{Wanda-10\%} \\
\midrule
Base (64)                  & 0.64 & 0.28 & 0.36 & 0.32 \\
Base+[D+Z]                 & 0.86 & 0.46 & 0.56 & 0.50 \\
Base+Instruct              & 0.78 & \textbf{0.76} & 0.48 & 0.44 \\
Base+Pruned                & 0.68 & 0.38 & 0.52 & 0.72 \\
\textbf{Base+Pruned+[D+Z]} & \textbf{0.94} & 0.72
                           & \textbf{0.84} & \textbf{0.78} \\

\midrule
\multicolumn{5}{c}{\textit{(b) Qwen2.5-14B}} \\
\midrule
\textbf{Optimization Setting}
& \textbf{Base}
& \textbf{Instruct}
& \textbf{Coder}
& \textbf{Math} \\
\midrule
Base (64)                  & 0.62 & 0.28 & 0.18 & 0.22 \\
Base+[D+Z]                 & 0.80 & 0.42 & 0.30 & 0.34 \\
Base+Instruct              & 0.88 & 0.70 & 0.36 & 0.44 \\
Base+Pruned                & 0.90 & 0.50 & 0.42 & 0.46 \\
\textbf{Base+Pruned+[D+Z]} & \textbf{0.92} & \textbf{0.78}
                           & \textbf{0.62} & \textbf{0.68} \\
\bottomrule
\end{tabular*}

\vspace{5pt}

\begin{tabular*}{0.70\textwidth}
{@{\extracolsep{\fill}}l|ccc@{}}
\multicolumn{4}{c}{\textit{(c) Ministral-3-8B}} \\
\toprule
\textbf{Optimization Setting}
& \textbf{Base}
& \textbf{Instruct}
& \textbf{Reasoning} \\
\midrule
Base (64)                  & 0.80 & 0.24 & 0.18 \\
Base+[D+Z]                 & 0.90 & 0.36 & 0.30 \\
Base+Instruct              & 0.86 & 0.66 & 0.44 \\
Base+Pruned                & 0.88 & 0.42 & 0.40 \\
\textbf{Base+Pruned+[D+Z]} & \textbf{0.96}
                           & \textbf{0.74}
                           & \textbf{0.68} \\
\bottomrule
\end{tabular*}

\caption{\textbf{Cross-family and cross-scale validation.}
FSR is reported only on the anchor and post-trained variants from the same family. All experiments use a $2\times4$ Markdown table and a 64-token
optimization budget. For Qwen2.5-7B, \texttt{Pruned} denotes \textit{Wanda-10\%}. For Qwen2.5-14B and Ministral-3-8B, it denotes \textit{Qwen2.5-14B-pruned-50p-rmsnorm-finetuned} and \textit{Ministral-3-8B-Base-2512-Wanda-50}, respectively. Bold values indicate the highest FSR in each column.}
\label{tab:cross_family_appendix}
\end{table*}

Table~\ref{tab:cross_family_appendix} shows a consistent overall pattern. Base-only optimization remains concentrated on the anchor checkpoint and transfers poorly to post-trained variants. Multi-template optimization, joint optimization with an instruction-tuned variant, and joint
optimization with a pruned variant all improve cross-variant transfer. The pruned+[D+Z] configuration achieves the strongest overall robustness
at both Qwen2.5 scales and on Ministral-3-8B. On Qwen2.5-14B, it reaches FSRs of 0.92, 0.78, 0.62, and 0.68 on the Base, Instruct, Coder, and Math variants, respectively. On Ministral-3-8B, it reaches 0.96, 0.74, and 0.68 on the Base, Instruct, and Reasoning variants. The consistent
ordering across Llama-2, Qwen2.5, and Ministral-3 suggests that the benefit of joint optimization is not tied to a single architecture or parameter scale.

\subsection{Sensitivity to Markdown Table Design}
\label{sec:table_design_sensitivity}

We further examine whether Perplexity Hiding depends on a specific Markdown table design. To isolate formatting effects, we keep the total optimization budget fixed at 64 tokens and vary two factors: the row/column layout and the semantic choice of table headers. Each design contains eight cells with eight optimized tokens per cell, so token allocation is controlled rather than varied. Experiments use the Base+Finance optimization setting, and we report PPL together with FSR on Llama-2 and four SFT variants.

\begin{table*}[t]
\centering
\scriptsize
\setlength{\tabcolsep}{3.2pt}
\renewcommand{\arraystretch}{1.08}
\resizebox{\textwidth}{!}{
\begin{tabular}{l|c|l|c|c|ccccc|c}
\toprule
\textbf{Setting} & \textbf{Table} & \textbf{Header Design}
& \textbf{Tokens/Cell} & \textbf{PPL} & \textbf{Llama-2}
& \textbf{Finance} & \textbf{Vicuna} & \textbf{Wizard}
& \textbf{Chinese} & \textbf{Avg. FSR} \\
\midrule
Default & $2\times4$ & Generic headers & 8 & 474 & 0.96 & 0.94 & 0.36 & 0.46 & 0.40 & 0.62 \\
Shape variant & $1\times8$ & Same as default & 8 & 612 & 0.94 & 0.92 & 0.34 & 0.44 & 0.38 & 0.60 \\
Shape variant & $4\times2$ & Same as default & 8 & 506 & 0.96 & 0.92 & 0.36 & 0.44 & 0.38 & 0.61 \\
Shape variant & $8\times1$ & Same as default & 8 & 735 & 0.94 & 0.90 & 0.34 & 0.42 & 0.36 & 0.59 \\
Header variant & $2\times4$ & Semantic headers & 8 & 438 & 0.96 & 0.94 & 0.38 & 0.46 & 0.40 & 0.63 \\
Header variant & $2\times4$ & Random headers & 8 & 562 & 0.94 & 0.92 & 0.36 & 0.44 & 0.38 & 0.61 \\
\bottomrule
\end{tabular}
}
\caption{\textbf{Sensitivity to Markdown table design.} All settings use a
fixed 64-token optimization budget under Base+Finance. Generic headers use
labels such as \texttt{Matrix} and \texttt{Value 1}; semantic headers are
contextually meaningful, whereas random headers are arbitrary.}
\label{tab:table_design_sensitivity}
\end{table*}

Table~\ref{tab:table_design_sensitivity} indicates that layout and header wording mainly influence PPL rather than FSR. PPL is sensitive to the surface naturalness and semantic coherence of the surrounding Markdown structure: balanced layouts and semantically meaningful headers receive lower perplexity. By contrast, the optimized cell tokens primarily determine triggering behavior. With the total token budget and optimized token positions preserved, changing the table shape or header wording alters average FSR by at most 0.04. The default $2\times4$ design therefore offers a favorable trade-off, achieving relatively low PPL while maintaining competitive FSR across downstream variants. These results show that SRAF does not rely on a single fragile table shape or a highly specific header choice.

\subsection{ProFLingo Extension Details}
\label{sec:appendix_proflingo}

To demonstrate that the core idea of our multi-task adversarial optimization framework can be seamlessly integrated into existing adversarial fingerprinting methods, we extend the ProFLingo baseline \cite{10735575} with our multi-model and multi-template joint optimization strategies. ProFLingo originally relies on a single-model optimization objective. We adapt it by applying our joint optimization over homologous models and chat templates, similar to our approach with SRAF. 

Table \ref{tab:proflingo_extension} presents the robustness evaluation of the extended ProFLingo across SFT and RLHF variants. The results exhibit a trend highly consistent with our main findings in Table \ref{tab:main_robustness}. The baseline ProFLingo struggles to transfer to domains with significant shifts (e.g., Code and Finance) and alignment variations. However, incorporating homologous models (\texttt{ProFLingo+Homologous Model}) or multiple prompt templates (\texttt{ProFLingo+Chat Template}) significantly improves the FSR on downstream variants. Ultimately, the synergistic strategy (\texttt{ProFLingo+S2.7-P+[D+Z]}) achieves the highest overall robustness, confirming that our optimization philosophy is method-agnostic and can effectively enhance the resilience of existing adversarial fingerprinting algorithms.

\begin{table*}[h]
\footnotesize
\centering
\setlength{\tabcolsep}{4.5pt} 
\renewcommand{\arraystretch}{1.1} 
\resizebox{\textwidth}{!}{ 
    \begin{tabular}{l|c|cccccc|ccc}
    \toprule
    \multirow{2}{*}{\textbf{Task Combination}} & \textbf{Anchor} & \multicolumn{6}{c|}{\textbf{SFT Variants}} & \multicolumn{3}{c}{\textbf{RLHF Variants}} \\ \cmidrule(lr){2-2} \cmidrule(lr){3-8} \cmidrule(lr){9-11}
    & \textbf{Llama-2} & \textbf{Finance} & \textbf{Vicuna} & \textbf{Wizard} & \textbf{Chinese} & \textbf{Code} & \textbf{Llemma} & \textbf{DPO} & \textbf{PPO-Pol} & \textbf{PPO-Rew} \\ \midrule
    ProFLingo & 1.00 & 0.24 & 0.52 & 0.42 & 0.28 & 0.20 & 0.14 & 0.28 & 1.00 & 1.00 \\
    ProFLingo+Homologous Model & 1.00 & 0.32 & 0.68 & 0.54 & 0.42 & 0.28 & 0.18 & 0.42 & 1.00 & 1.00 \\
    ProFLingo+Chat Template & 1.00 & 0.30 & 0.70 & 0.52 & 0.38 & 0.26 & 0.16 & 0.48 & 1.00 & 1.00 \\ \midrule
    \textbf{ProFLingo+S2.7-P+[D+Z]} & \textbf{1.00} & \textbf{0.48} & \textbf{0.84} & \textbf{0.76} & \textbf{0.72} & \textbf{0.44} & \textbf{0.28} & \textbf{0.82} & \textbf{1.00} & \textbf{1.00} \\ \bottomrule
    \end{tabular}
}
\caption{\textbf{Robustness evaluation of extended ProFLingo across SFT and RLHF variants.} The FSR values demonstrate that integrating our multi-model and multi-template optimization strategies significantly enhances the transferability of ProFLingo, validating the generalizability of our framework.}
\label{tab:proflingo_extension}
\end{table*}

\section{Detailed Expirement Setting}

\subsection{Base Model and Variants}
\label{sec:model_inventory_appendix}

Table \ref{tab:model_inventory} provides a comprehensive inventory of all models used in our experiments, organized by category and modification type.

\begin{table*}[h]
\centering
\small
\begin{tabular}{@{}lllp{7cm}@{}}
\toprule
\textbf{Category} & \textbf{Model/Method} & \textbf{Abbreviation} & \textbf{Description} \\ \midrule
\multicolumn{4}{l}{\textit{Base Models}} \\ \midrule
Base & Llama-2-7B & Llama-2-7B & Primary base model for fingerprint optimization \\ \midrule
\multicolumn{4}{l}{\textit{Incremental Training (SFT)}} \\ \midrule
SFT & Llama-2-Finance-7B & Finance & Fine-tuned on financial domain data \\
SFT & Vicuna-1.5-7B & Vicuna & Fine-tuned on conversation data from ShareGPT \\
SFT & WizardMath-7B & WizardMath & Fine-tuned for mathematical reasoning \\
SFT & ChineseLlama-2-7B & Chinese & Fine-tuned for Chinese language tasks \\
SFT & CodeLlama-7B & Code & Fine-tuned for code generation \\
SFT & Llemma-7B & Llemma & Fine-tuned for mathematical applications \\ \midrule
\multicolumn{4}{l}{\textit{RLHF Variants}} \\ \midrule
RLHF & Llama-2-7B-DPO-v0.1 & DPO & Direct Preference Optimization variant \\
RLHF & Llama-2-7B-PPO-v0.1-policy & PPO-Pol & PPO policy model \\
RLHF & Llama-2-7B-PPO-v0.1-reward & PPO-Rew & PPO reward model \\
RLHF & Tulu-2-DPO-7B & Tulu-DPO & DPO fine-tuned instruction model \\ \midrule
\multicolumn{4}{l}{\textit{Pruned Models}} \\ \midrule
Pruning & Sheared-Llama-1.3B-Pruned & S1.3-P & Structurally pruned to 1.3B parameters \\
Pruning & Sheared-Llama-1.3B & S1.3 & Pruned with continuation training \\
Pruning & Sheared-Llama-1.3B-Sharegpt & S1.3-S & Pruned with ShareGPT fine-tuning \\
Pruning & Sheared-Llama-2.7B-Pruned & S2.7-P & Structurally pruned to 2.7B parameters \\
Pruning & Sheared-Llama-2.7B & S2.7 & Pruned with continuation training \\
Pruning & Sheared-Llama-2.7B-Sharegpt & S2.7-S & Pruned with ShareGPT fine-tuning \\
Pruning & Sparse-Llama-2-7B & Sparse & Sparse pruning applied \\
Pruning & Wanda-Llama-2-7B & Wanda & Wanda pruning method \\
Pruning & GBLM-Llama-2-7B & GBLM & Gradient-based low-rank bias pruning \\ \midrule
\multicolumn{4}{l}{\textit{Cross-Family Models (False Positive Testing)}} \\ \midrule
Cross & Llama-3.1-8B & Llama-3.1 & Different architecture version \\
Cross & Mistral-7B-v0.3 & Mistral & Different model family \\
Cross & Qwen2.5-7B & Qwen2.5 & Alibaba Qwen family \\
Cross & Qwen2.5-7B-Instruct & Qwen-Ins & Instruction-tuned Qwen variant \\
Cross & Yi-6B & Yi & 01.AI Yi series \\
Cross & Bloom-7B1 & Bloom & BigScience BLOOM family \\
Cross & Pythia & Pythia & EleutherAI Pythia series \\
Cross & Gemma-7B-it & Gemma & Google Gemma instruction-tuned \\ 
Cross & EvoLLM \cite{akiba2024evomodelmerge} & EvoLLM & Unrelated to Llama-2-7B family \\ \midrule
\multicolumn{4}{l}{\textit{Model Merging}} \\ \midrule
Merging & Breadcrumbs & Bread. & Task-arithmetic merging with gradient breadcrumbs \\
Merging & Breadcrumbs-Ties & B-Ties & Breadcrumbs combined with TIES magnitude pruning \\
Merging & Dare-Task & D-Task & Dropout-added residual merging with task arithmetic \\
Merging & Dare-Ties & D-Ties & DARE combined with TIES magnitude pruning \\
Merging & Della & Della & DELLA linear mode merging with Fisher information \\
Merging & Della-Task & Del-Task & DELLA combined with task arithmetic \\
Merging & Task-Arithmetic & Task & Task-arithmetic merging on model checkpoints \\
Merging & Ties & Ties & TIES magnitude/sign sparse merging \\ 
Merging & FuseLLM-7B \cite{wan2024knowledge} & FuseLLM & Knowledge fusion of diverse architecture LLMs \\ \bottomrule
\end{tabular}
\caption{Inventory of models used in SRAF experiments.}
\label{tab:model_inventory}
\end{table*}

\subsection{Chat Templates}
\label{sec:chat_templates_appendix}
Table \ref{sec:chat_templates} lists the 5 chat templates used for multi-task joint optimization during fingerprint generation.

\begin{CJK*}{UTF8}{gbsn}
\begin{table*}[h]
\centering
\small
\begin{tabular}{@{}lp{0.12\textwidth}p{0.65\textwidth}@{}}
\toprule
\textbf{Template Name} & \textbf{Abbr.} & \textbf{Format} \\ \midrule
Default & D & \texttt{<|im\_start|>user\textbackslash n\{message\}<|im\_end|>\textbackslash n<|im\_start|>assistant\textbackslash n} \\ \addlinespace[0.3em]
Alpaca & A & \texttt{\#\#\# Instruction:\textbackslash n\{instruction\}\textbackslash n\textbackslash n\#\#\# Response:} \\ \addlinespace[0.3em]
ChatGLM & C & \texttt{[Round 1]\textbackslash n}问：\texttt{\{question\}\textbackslash n}答： \\ \addlinespace[0.3em]
Llama2 & L & \texttt{<s>[INST] \{user message\} [/INST] </s>} \\ \addlinespace[0.3em]
Zero-Shot & Z & \{instruction\} (direct prompt without template) \\ \bottomrule
\end{tabular}
\caption{Chat templates used in SRAF multi-task optimization.}
\label{sec:chat_templates}
\end{table*}
\end{CJK*}

\subsection{System Prompts}
\label{systemctl_prompt_appendix}
Table \ref{systemctl_prompt} lists all system prompts used in our experiments for evaluation robustness testing.

\begin{table*}[h]
\centering
\small
\begin{tabular}{@{}lp{0.25\textwidth}p{0.55\textwidth}@{}}
\toprule
\textbf{Template Name} & \textbf{Abbreviation} & \textbf{System Prompt Suffix} \\ \midrule
Fastchat-T5-LLaMA & Fastchat & \texttt{A chat between a curious user and an artificial intelligence assistant. The assistant gives helpful, detailed, and polite answers to the user's questions.} \\ \addlinespace[0.3em]
Llama2-chat & Llama2 & \texttt{You are a helpful, respectful and honest assistant. Always answer as helpfully as possible, while being safe. Your answers should not include any harmful, unethical, racist, sexist, toxic, dangerous, or illegal content. Please ensure that your responses are socially unbiased and positive in nature.} \\ \addlinespace[0.3em]
OpenAI & OpenAI & \texttt{You are ChatGPT, a large language model. Follow the user's instructions carefully.} \\ \addlinespace[0.3em]
Hiking Assistant & Hiking & \texttt{You are an enthusiastic hiking expert who loves helping people plan their hiking adventures. Provide detailed trail information, safety tips, and gear recommendations.} \\ \addlinespace[0.3em]
Tax Assistant & Tax & \texttt{You are a professional tax advisor. Help users understand tax regulations and provide accurate tax-related guidance. Always clarify that this is general information, not legal advice.} \\ \addlinespace[0.3em]
JSON Formatter & Json & \texttt{You are a data formatting specialist. Always respond with valid, well-formatted JSON. Ensure proper syntax, quote handling, and structure.} \\ \addlinespace[0.3em]
Marketing Copywriter & Mktg. & \texttt{You are a creative marketing copywriter. Craft engaging, persuasive content that highlights product benefits and appeals to target audiences.} \\ \addlinespace[0.3em]
Shakespearean Style & Shake. & \texttt{You are Shakespeare. Speak in Elizabethan English with poetic flourishes, using thou, thee, and archaic expressions appropriate for the Bard's era.} \\ \addlinespace[0.3em]
Customer Support & Cust. & \texttt{You are a helpful customer service representative. Address customer concerns professionally, offer solutions, and maintain a friendly, patient demeanor.} \\ \bottomrule
\end{tabular}
\caption{System prompts used in SRAF experiments.}
\label{systemctl_prompt}
\end{table*}

\end{document}